\documentclass[traditabstract]{aa}

\usepackage{graphicx}
\usepackage{natbib}
\usepackage{lscape}
\usepackage{supertabular}
\usepackage{xcolor}

\newcommand{\itio}[0]{I$_{\mathrm{TiO}}$}

\newcommand{\cairta}[0]{\ion{Ca}{ii}\,IRT$_{\rm 1}$}
\newcommand{\cairtb}[0]{\ion{Ca}{ii}\,IRT$_{\rm 2}$}
\newcommand{\cairt}[0]{\ion{Ca}{ii}\,IRT}
\newcommand{\cahk}[0]{\ion{Ca}{ii}~H \& K\,}
\newcommand{\rpr}[0]{R$^{\prime}_{\rm HK}$\,}

\usepackage{txfonts}

\begin{document}

\title{The CARMENES search for exoplanets around M dwarfs}
\subtitle{Variability on long timescales as seen in chromospheric indicators}

\author{B. Fuhrmeister\inst{\ref{inst1}}, S. Czesla\inst{\ref{inst13},\ref{inst1}},  V. Perdelwitz\inst{\ref{inst1a},\ref{inst1}}
  \and  E. Nagel\inst{\ref{inst1}}
  \and J. H. M. M. Schmitt\inst{\ref{inst1}}
  \and  S.~V.~Jeffers\inst{\ref{inst15}}
  \and  J.~A.~Caballero\inst{\ref{inst3}}
  \and M.~Zechmeister\inst{\ref{inst2}}
  \and  D.~Montes\inst{\ref{inst10}}
  \and   A.~Reiners\inst{\ref{inst2}}
  \and  \'A.~L\'opez-Gallifa \inst{\ref{inst10}}
  \and   I.~Ribas\inst{\ref{inst5},\ref{inst4}}
  \and  A.~Quirrenbach\inst{\ref{inst7}}
  \and P.~J.~Amado\inst{\ref{inst6}}
  \and  D.~Galad\'{\i}-Enr\'{\i}quez\inst{\ref{inst12}}
  \and  V.~J.~S.~B\'ejar\inst{\ref{inst8},\ref{inst9}}
  \and  C.~Danielski\inst{\ref{inst6}}
  \and  A.~P.~Hatzes\inst{\ref{inst13}}
  \and  A. Kaminski\inst{\ref{inst7}}
  \and  M.~K\"urster\inst{\ref{inst14}}
  \and  J.~C.~Morales\inst{\ref{inst5},\ref{inst4}}
  \and  M.~R.~Zapatero Osorio\inst{\ref{inst16}}}


\institute{Hamburger Sternwarte, Universit\"at Hamburg, Gojenbergsweg 112, D-21029 Hamburg, Germany\\
  \email{bfuhrmeister@hs.uni-hamburg.de}\label{inst1}
        \and
     Th\"uringer Landessternwarte Tautenburg, Sternwarte 5, D-07778 Tautenburg, Germany\label{inst13} 
        \and
        Kimmel fellow, Department of Earth and Planetary Sciences, Weizmann Institute of Science, Rehovot, 76100, Israel \label{inst1a} 
        \and
        Max-Planck-Institut f\"ur Sonnensystemforschung, Justus-von-Liebig-Weg 3,37077 G\"ottingen, Gemany\label{inst15}
        \and
        Centro de Astrobiolog\'{\i}a (CAB), CSIC-INTA, Camino Bajo del Castillo s/n, E-28692 Villanueva de la Ca\~nada, Madrid, Spain \label{inst3}
        \and
        Institut f\"ur Astrophysik, Friedrich-Hund-Platz 1, D-37077 G\"ottingen, Germany\label{inst2} 
        \and
        Facultad de Ciencias F\'{\i}sicas, Departamento de F\'{\i}sica de la Tierra y Astrof\'{\i}sica; IPARCOS-UCM (Instituto de F\'{\i}sica de Part\'{\i}culas y del Cosmos de la UCM), Universidad Complutense de Madrid, E-28040 Madrid, Spain\label{inst10} 
        \and
        Institut d'Estudis Espacials de Catalunya, E-08034 Barcelona, Spain\label{inst5}
        \and  
        Institut de Ci\`encies de l'Espai (CSIC), Campus UAB, c/ de Can Magrans s/n, E-08193 Bellaterra, Barcelona, Spain\label{inst4}
        \and
        Landessternwarte, Zentrum f\"ur Astronomie der Universit\"at Heidelberg, K\"onigstuhl 12, D-69117 Heidelberg, Germany\label{inst7} 
        \and
        Instituto de Astrof\'isica de Andaluc\'ia (CSIC), Glorieta de la Astronom\'ia s/n, E-18008 Granada, Spain\label{inst6} 
        \and
        Centro Astron\'omico Hispano en Andaluc\'ia, Observatorio Astron\'omico de Calar Alto, Sierra de los Filabres, E-04550 G\'ergal, Almer\'{\i}a, Spain\label{inst12} 
        \and
        Instituto de Astrof\'{\i}sica de Canarias, c/ V\'{\i}a L\'actea s/n, E-38205 La Laguna, Tenerife, Spain\label{inst8}
        \and
        Departamento de Astrof\'{\i}sica, Universidad de La Laguna, E-38206 Tenerife, Spain\label{inst9} 
        \and
        Max-Planck-Institut f\"ur Astronomie, K\"onigstuhl 17, D-69117 Heidelberg, Germany\label{inst14}
        \and
       Centro de Astrobiolog\'{\i}a (CAB), CSIC-INTA, Carretera de Ajalvir, km~4, E-28850
Torrej\'on de Ardoz, Madrid, Spain\label{inst16}
       }
        
\date{Received 29/08/2022; accepted dd/mm/2022}

\abstract{ It is clearly established that the Sun has an  11-year cycle that is caused by its
internal magnetic field.
  This cycle is also observed in a sample of M dwarfs. In the framework of exoplanet detection
  or atmospheric characterisation of exoplanets, the activity status of the
  host star plays a crucial role, and inactive states are preferable for such studies. This means that  it is important to know the activity cycles of these stars. We study systematic long-term variability
  in a sample of 211 M dwarfs observed with CARMENES, the high-resolution optical and near-infrared
  spectrograph at Calar Alto Observatory. In an automatic search using time series of different activity indicators, we 
  identified 26 stars with linear or quadratic trends or with potentially cyclic behaviour. 
   Additionally, we performed an independent search in archival \rpr\ data collected from different
    instruments whose time
    baselines were usually much longer. These data are available for a subset of 186 of our sample stars. Our search
  revealed 22 cycle candidates in the data.
  We found that
  the percentage of stars showing long-term variations drops dramatically to the latest  M dwarfs.
  Moreover, we found that the pseudo-equivalent width (pEW) of the  H$\alpha$ and \ion{Ca}{ii} infrared triplet
   more often triggers automatic detections of long-term variations 
  than the TiO index,  differential line width,  chromatic index, or radial velocity. This is in line with
  our comparison of the median relative amplitudes of the different indicators. For stars that trigger our automatic detection,
  this leads to the highest amplitude variation in \rpr\, , followed by pEW(H$\alpha$), pEW(\cairt), and the TiO index.
}
    
\titlerunning{Long-term variability in chromospheric indicators of CARMENES stars}
\authorrunning{B. Fuhrmeister et~al.}
\maketitle


\section{Introduction}

Stellar activity is driven by an underlying magnetic dynamo that  is
predicted to change periodically according to the theory of an $\alpha$-$\Omega$ dynamo \citep{Roberts1972}.
Dynamo simulations for the fully convective M dwarf Proxima Centauri also directly led
to magnetic cycles  \citep{Yadav2016}. 
This periodic behaviour manifests itself as stellar activity cycles. These cycles have also been
observed on stars other than the Sun by time series of X-ray and chromospheric activity indicators, or
  photometry. 
While the Sun varies in X-rays as a coronal indicator by more than a factor of ten during
its cycle \citep{Peres2000}, only a few stars have been found to exhibit long-term cyclic behaviour in X-rays. These cycles all had a much lower amplitude \citep{Robrade2012, Wargelin2017, Coffaro2020}.

Chromospheric indicators, on the other hand, have been used widely for a cycle search
in solar-type stars. The monitoring program conducted at the Mt. Wilson Observatory
\citep{Wilson1978} using the line index of the \cahk lines revealed that 
stellar activity cycles are a ubiquitous
phenomenon \citep{Baliunas1995}. Short and long
cycles have been found for several stars \citep{Brandenburg2017}, starting with the Sun,
which exhibits the approximately 80-year-long
Gleissberg cycle \citep{Gleissberg1945} and the 11-year-long Schwabe cycle 
\citep{Richards2009}, although it also shows a
transient periodic behaviour of about one-half to two years \citep{Ballester2002, Mursula2003}.  Multiple cycles have also been found in other stars, such as 
in $\epsilon$ Eridani, which displays  two distinct cycles at about three and twelve years \citep{Brandenburg2017, Jeffers2022}.
In addition to these decade-long cycles, F-type stars with activity cycles
shorter than one year have been revealed \citep{Jeffers2018a, Mittag2019}.
Although some
authors sort  different cycles lengths into an inactive and active branch,
the existence of these two branches has been questioned by \citet{BoroSaikia2018} and
\citet{Brown2022}, for example.

While extensive work about activity cycles has been conducted on FGK stars (see e.g.
\citet{Lovis2011}), only a few M dwarfs
have been under investigation, mostly using photometric time series.
For example, \citet{Suarez-Mascareno2016} derived magnetic cycles and rotation periods in 125 late-type
stars, 70 of which were M dwarfs. About half of them exhibited long-term
photometric variability. Another study that used photometric data to derive
rotation periods and cycles was conducted by \citet{DA19}. They found 12 stars with long-term cyclic variations with an observation
time baseline of more than nine years. 
Moreover, \citet{Kueker2019} found that in
a sample of 31 M dwarfs, 19 exhibited a cycle. Using chromospheric indicators, fewer cycles were found for M dwarfs by
\citet{Robertson2013} by employing time series of H$\alpha$ indices. They also identified
linear and quadratic trends as possible parts of activity cycles, which are too long
to be covered by the available data. While they searched 93 K and M dwarfs, only six periods and
seven long-term trends were identified  by \citet{Robertson2013}. Using H$\alpha$ and \ion{Ca}{ii} H\&K data,
  \citet{SuarezMascareno2018}
found that in a sample of 71 M dwarfs, 13 exhibited a cycle. Moreover, \citet{GdS12}
found 
a correlation between
long-term activity variations measured by a \ion{Na}{i} D index and radial velocity (RV) in about one-third of the 27 stars they analysed.

This long-term correlation between an activity indicator and RV may impose problems
for exoplanet detections with longer orbital periods, as is already known from short
time-activity variations caused by the rotation period of the star, for instance, which
may hide or mimic a planetary RV signal  \citep{Queloz2001,Dumusque2011,Jeffers2022a}. This
is caused by the asymmetry that is imprinted on the spectrum by surface heterogeneities. The
asymmetry in turn leads to a shift of the line centre depending on the size
and location of the active region on the stellar surface. 
Accordingly, \citet{SN20} found a long-term periodic behaviour in their RV
measurements, which were mainly conducted with the CARMENES spectrograph. In the framework
of an exoplanet search, they found
tentative activity cycles of 600 and 2900\,d for the two M dwarfs GJ~251 and Lalande~21185, respectively.

This study searches for more activity cycles in M dwarfs using data from the
CARMENES spectrograph for different chromospheric line measurements.
The paper is structured as follows: In Sect. \ref{sec:obs} we present the data, their reduction,
and the method for the measurement of our chromospheric indicators. In Sect. \ref{sec:methods} we
detail our search methods for  activity cycles and trends.
In Sect. \ref{sec:results} we present and discuss our results.  We
give a summary and concluding remarks in Sect. \ref{conclusion}.

\section{Observations and measurements of chromospheric indicators}
\label{sec:obs}

All spectra used for the present analysis were taken
with the CARMENES spectrograph, installed at the 3.5\,m Calar Alto 
telescope \citep{Quirrenbach2020}.
CARMENES is a spectrograph covering the wavelength range
from 5200 to 9600\,\AA\, in the visual channel (VIS) and from 9600 to 17\,100\,\AA\, in
the near-infrared
channel (NIR). The instrument provides a spectral resolution of
$\sim$ 94\,600 in VIS and $\sim$ 80\,400 in NIR. 
The CARMENES consortium is conducting a 
survey of $\sim$350 M~dwarfs to find low-mass exoplanets \citep{AF15a, Reiners2017}.
Since the cadence of the spectra is optimised for the planet search, usually
no continuous time series are obtained.

\subsection{Data reduction and sample construction}
\begin{figure}
\begin{center}
\includegraphics[width=0.5\textwidth, clip]{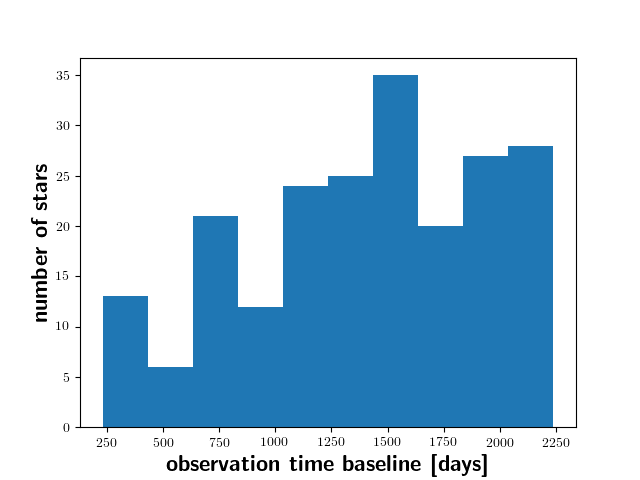}\\
\caption{\label{baseline} Distribution of the observation time baseline
  for our sample of 211 stars with more than 19 observations and a minimum observation
  time baseline of 200 days.
}
\end{center}
\end{figure}

In our analysis, we considered a
sample of 362 M~dwarfs observed by CARMENES, resulting in  
more than 19\,000 spectra taken before February 2022. From this sample, we excluded known binaries
\citep{Baroch2018,Schweitzer2019,Baroch2021}. Binaries may
hamper our analysis because the line shifts induced by binarity may be large enough
to shift parts of the line out of the integration range for line indices or pseudo-equivalent width (pEW).
Additionally, we had to constrain the stellar sample further because the CARMENES data are optimised for exoplanetary search, and therefore not all stars are
observed with the same frequency or time baseline (i.e. the elapsed time between the first and last
observation). We wish to search for
long-term systematic variations, therefore we restricted the sample to all stars with a time coverage
of more than 200 days with at least 20 observations to allow for a statistically
meaningful analysis. Even though it is hard to detect periods with so few observations,
linear and quadratic trends may still be detected. In this fashion, we obtained
a sub-sample of 211
stars for our analysis.
We show the distribution of the observation time baseline for these 211 stars in Fig. \ref{baseline}.

The stellar spectra
were reduced using the CARMENES reduction pipeline
\citep{pipeline,Caballero2}. Subsequently, we corrected them for barycentric and 
RV
motions and carried out a correction for telluric absorption lines \citep{Evangelos} 
using the {\tt molecfit}
package\footnote{\tt{https://www.eso.org/sci/software/pipelines/skytools /molecfit}}.
No correction for airglow emission lines was attempted because they do not play a role for the activity indicators we used.

\subsection{Measurement of activity indicators}

To assess the activity state of the stars in each spectrum, we employed pEW measurements
because the spectra of M dwarfs do not show an identifiable continuum (because molecular
absorption lines are present nearly everywhere).
We measured the pEW of the blue \ion{Ca}{ii} infrared triplet (IRT) line at 8500.35\,\AA\, 
(hereafter \cairta) and of the middle \ion{Ca}{ii} IRT  line at 8544.44\,\AA\, (hereafter \cairtb), and 
the H$\alpha$ line (all wavelengths are given for vacuum conditions). 
The three \cairt\, lines act very similarly regarding activity \citep{Lafarga2021},
therefore we used only the two bluer lines because they are located on the same order as the spectrograph.

The pEW was calculated using the expression
\begin{equation}
        \mathrm{pEW}=\int (1- F_{\rm core}/F_{0})d\lambda,
\end{equation}  
where $F_{core}$ is the flux density in the line band, $F_{0}$ is the mean flux density in the two reference intervals (representing the pseudo-continuum), and $\lambda$ denotes the wavelength.
Thus, time series of pEW measurements were obtained for each star. The  integration
ranges for the individual lines are given in Table~\ref{ew}.

The uncertainty of individual
pEW measurements  depends on the S/N of the spectra, which can be particularly problematic
for the H$\alpha$ line, where the S/N is typically better
for bright, early M~dwarfs than for fainter mid-type M dwarf. 
We did not compute pEWs when the S/N in the pseudo-continuum was
lower than 15. Therefore, no pEW(H$\alpha$) measurements
could be obtained for some of the latest and faintest  M~dwarfs,  and we omitted these stars
from further analysis when fewer than 20  measurements were available. When the S/N
was good enough to compute pEWs, we calculated their error by a bootstrap analysis
using the flux density errors from the pipeline to re-compute the pEWs 1000 times.
This yielded quite low statistical
errors;  in the case of pEW(H$\alpha$) and pEW(\cairt), the relative statistical
  error is about $10^{-7}$--$10^{-5}$ for most of the spectra, while the time series of the chromospheric
indicators of inactive stars exhibits a larger scatter of about 10\% of the median
value of the pEW in H$\alpha$ and of 3\% in both pEW(\cairt). This is apparently caused by intrinsic variations of the
activity level of the stars, and we therefore applied this as error to all chromospheric
indicator measurements. We also display this scatter as error bars in all the figures 
showing time series.

In addition to these classical chromospheric indicators, we also considered more recently
introduced activity indicators, specifically, a TiO index defined as the ratio of the average flux
density in a band red- and blue-wards of the band head, as introduced by \citet{Patrick}. The
wavelength bands, which slightly differ from \citet{Patrick},  are given in Table~\ref{ew}.
An error was again estimated by the variation in inactive stars to be 0.3\% (which roughly
agrees with a bootstrap analysis in this case).
Moreover, we used RV, chromatic index (CRX; as a measure of change in RV shift
as a function of wavelength), and the differential line width
(dLw; as a measure of changes in line width compared to a constructed average).
All three indicators and their errors were obtained for the VIS arm of CARMENES from the reduction
pipeline \texttt{serval} \citep{serval}.
They have been used for example in a search for rotation periods by \citet{Lafarga2021}.

\begin{table}
        \caption{\label{ew} Parameters (vacuum wavelength) of the pEW and TiO index calculation. }
\footnotesize
\begin{tabular}[h!]{lcccc}
\hline
\hline
\noalign{\smallskip}

Line           & Wave-   & Width  & Reference  & Reference  \\
& length      &       & band 1 &  band 2 \\
& [\AA] & [\AA] & [\AA] & [\AA]\\
\noalign{\smallskip}
\hline
\noalign{\smallskip}
H$\alpha$& 6564.60 & 1.60 & 6537.4--6547.9 & 6577.9--6586.4 \\
\cairta    & 8500.35 & 0.50 & 8476.3--8486.3 & 8552.4--8554.4 \\
\cairtb    & 8544.44 & 0.50 & 8476.3--8486.3 & 8552.4--8554.4 \\
TiO       &          &      & 7056.2--7059.2   & 7046.0--7050.0\\
\noalign{\smallskip}
\hline

\end{tabular}
\normalsize
\end{table}

For a sub-sample of 186 stars  with \rpr
measurements compiled by \citet{Perdelwitz2021}
from seven different instruments (most spectra were obtained with the HARPS\footnote{High Accuracy Radial velocity Planet Searcher},
HIRES\footnote{High Resolution Echelle Spectrometer}, and Narval spectrographs), we included these data in our cycle search.
\citet{Perdelwitz2021} used the pipeline-reduced spectra
from the respective archives and computed the chromospheric line fluxes in the \cahk lines after
rectification and subtraction of the photospheric flux using PHOENIX \citep{Hauschildt1999} photospheric model spectra 
from the \citet{Husser2013} database. While some of the obtained \rpr time series temporally overlap with the CARMENES
data, most of them do not. Moreover, the observational cadence is usually coarser, while the time baseline
is much longer than for the CARMENES data. This longer baseline makes the data more suitable for a cycle search.

\subsection{Compilation of cycles from the literature}
From the literature, we collected a list of 57 proposed cycles in 41 stars as
linear or quadratic trends for our sample stars. Most of these cycles
originate from an analysis of photometric data. For example, the studies by \citet{Suarez-Mascareno2016},
which were dedicated to a cycle search, and of \citet{DA19}, where some cycles emerged as by-products
of a rotation period search, used only photometric data. We also included cycles found
in photometric data by \citet{Kueker2019}.
In contrast, \citet{Robertson2013} used an H$\alpha$ index for an activity cycle search,
and \citet{Perdelwitz2021} also performed a period analysis on their \rpr data with
three published cycles. We also have some stars in common with the sample
of \citet{SuarezMascareno2018}, who used the \ion{Ca}{ii} H\&K and H$\alpha$ index
for their cycle search. 

Other activity cycles have been published as part of RV exoplanet searches
that also used the CARMENES data, for example by \citet{SN20} for Lalande~21185 (including
RV data from the SOPHIE instrument) and GJ~251 (including data from the HARPS instrument),
for which cycle lengths of 2900 days and 600 days were found. Moreover, \citet{Lopez-Santiago2020} found a probable cycle length of 14 years  for the star 
GJ~3512 by performing a combined fit to the
rotation period and activity cycle. 

\section{Searching for systematic long-term variations}\label{sec:methods}

We used several methods to search for systematic long-term variations in the available data to
account for the different time scales involved.
While we searched for cycles that were potentially longer than a decade, our longest
time baseline is only 6.1\,a (= years). This implies that we may cover only parts of activity
cycles. Therefore, we also searched for linear and quadratic trends, which may arise when the data cover only the gradual decay or increase in activity or
include an activity maximum or minimum. We caution, nevertheless, that these
systematic long-term variations need not to lead to periodic behaviour. Therefore,
we distinguish between periodic behaviour, where we do cover at least one full period,
and linear and quadratic trends, which are indicative of systematic, but not necessarily
cyclic behaviour.


Moreover, in all our search algorithms, we decided to tie a detection to at least three
indicators and  used in turn not as strict criteria for the individual indicators, as
we would have applied if we had
accepted detections for single indicators.  We
opted for this alternative since correlated (or anti-correlated)
behaviour between the indicators is expected in the case of systematic long-term 
variability since they (optimally) all should react to the changing activity level.
 We decided in this context to use at
  least three indicators, because for only two indicators, the two used \cairt\ lines may
  trigger a detection alone. With more than three indicators, this criterion is very
strict because it may lead to non-detections when the different 
indicators have different sensitivities. 

Finally, before searching for cycles or parts of it, we applied a 3$\sigma$ clipping to each
of the indicator time series using the {\tt python} routine {\tt scipy.sigmaclip}. 
Thus we avoided
outliers caused by flaring, bad weather, or instrumental effects, which may 
confound the search. 
Moreover, after conducting the automatic variability search, we excluded all time series that had data gaps spanning more than 40\%
of the temporal baseline, were longer than 550 days, or both, because we
noted visually that long data gaps like this were often problematic.

\subsection{Searching for linear trends}\label{sec:trends}
The linear trend search was conducted as follows: We used each of the pEW of H$\alpha$, \cairta, \cairtb,
and the \itio\, indices as well as CRX and dLw to compute a linear regression with
respect to time,
employing {\tt scipy.stats.linregress} in {\tt python,} which outputs also
Pearson's correlation coefficient $r$
and the corresponding $p$-value. This additional output
allowed us to identify trends in time.  These trends were searched for using the whole data set for
  each star because we are interested in finding stars for which the whole time series covers
  the activity increase or decrease phase of an activity cycle. The identified trends
  are linear by the definition
of the Pearson's correlation coefficient, but a subsequent visual inspection showed
that in a few cases, the linear description only holds as a first-order
approximation, but a parabolic description appears to be more
appropriate. These time series were also found in our quadratic trend search and were marked accordingly. 
Therefore, we claim a linear trend when the Pearson
correlation coefficients are
abs$(r)>0.55$ and  $p<0.002$ in three or more of the indicators.
 We determined these thresholds empirically using a test sample
of 20 stars that included 3 stars whose linear trends were found by eye. 
The relatively low threshold of abs$(r)>0.55$ was justified a posteriori because we did
not
reject any of the automatically found trends by eye (but only found some to be
better described by quadratic trends).  

\begin{figure}
\begin{center}
\includegraphics[width=0.5\textwidth, clip]{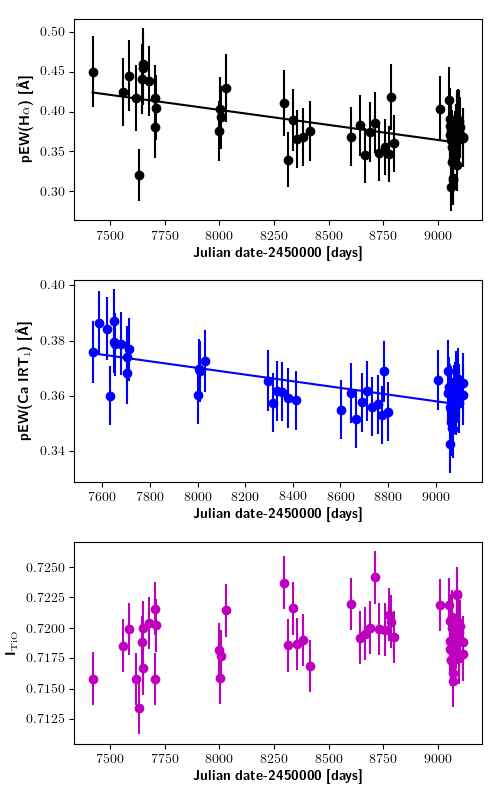}\\
        \caption{\label{fig:trend} Time series of J23245+578 / BD+57~2735. 
        We show the pEW(H$\alpha$) (\emph{top}, black dots), 
        pEW(\cairta) (\emph{middle}, blue dots), and \itio\, (\emph{bottom}, magenta dots).
        Additionally, the best linear fit 
        is indicated as the solid line.  The error bars are not statistical,
but determined by the activity jitter of inactive stars.
}
\end{center}
\end{figure}
\begin{figure}
\begin{center}
\includegraphics[width=0.5\textwidth, clip]{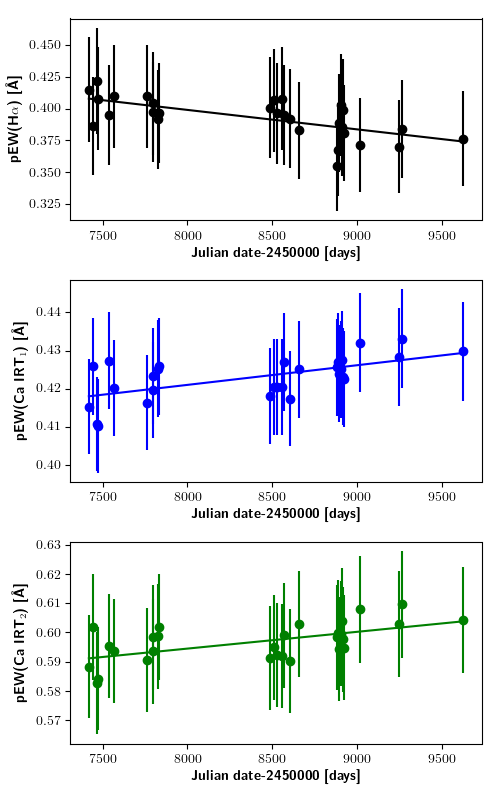}\\
        \caption{\label{fig:trend2} Time series of J13450+176 / BD+18~2776.
        We show the pEW(H$\alpha$) (\emph{top}, black dots),
         pEW(\cairta) (\emph{middle}, blue dots), and pEW(\cairtb) (\emph{bottom}, green dots). Additionally, the best-fit linear trend
         is indicated as the solid line.
}
\end{center}
\end{figure}

An instructive example of the trends 
we found is shown in Fig.~\ref{fig:trend} for the M1.0\,V star J23245+578 / BD+57~2735.
The pEW(H$\alpha$) and pEW(\cairta) both show about the same slope. This
is the case for most of the trends we found. In one of the time series,
we nevertheless found one convincing example in which the trends of H$\alpha$ and the pEW(\cairta) were of opposite sign. 
In the M0.0\,V star J13450+176 / BD+18~2776,
we found a highly significant Pearson correlation coefficient of 0.95 ($p$-value $10^{-15}$) between the  the values of  pEW(\cairta) and pEW(\cairtb), 
while the  correlation of  the values of pEW(H$\alpha$) and the two pEW(\cairt) are
$-0.73$ and $-0.69$, respectively ($p$-values $< 3.1\cdot 10^{-5}$). We show the corresponding
time series in Fig. \ref{fig:trend2} with the pEW(\cairtb)
instead of \itio\ as an exception.
This behaviour may be explained by the 
relative inactivity of the star. With an increase in activity level in M dwarfs, we first expect
a deepening of the H$\alpha$ line, which is followed by
a fill-in, until the line finally enters emission at even higher activity 
levels \citep{Cram1979}. For the \cairt\ lines, only a fill-in is expected instead. This
behaviour can lead to an anti-correlation of the two lines for the lowest activity stars, where
H$\alpha$ still deepens with increasing activity.

\subsection{Searching for quadratic trends}\label{sec:maxima}
We used
a parabolic fit
to identify another type of systematic long-term variability.
We applied a collection
of ten criteria, nine of which had to be fulfilled to accept a quadratic trend.
We used a reduced $\chi^{2}$ like quantity to access the quality of the fit.
However, the correspondence is not exact owing
to our use of a generic jitter estimate and the absence of the degrees of
freedom in the definition:
(i) $\chi^{2} = \frac{1}{n_{\mathrm{spec}}}\frac{\sum (pEW_{i}-poly)^{2}}{\sum err_{i}^{2}} <  1.05$ for pEW(H$\alpha$) (ii-iii) The same holds for \cairta\, and \cairtb\,
with
a threshold of 1.0 (iv) The same holds for the TiO
index with a threshold of 1.03. 
To fine-tune these first four criteria, we visually inspected a sub-sample of 20 time series,  which
included five time series for which a quadratic trend was found by eye. 
(v)--(viii) All  maxima or minima of the parabolic fit had to lie within the 
observed time baseline. (ix)  The maximum difference between the
corresponding dates of the maxima or minima of the quadratic fits of the different
activity indicators
must be lower than 15\% of the 
observation time baseline and shorter than 150 days, with the exception of one indicator.
(x) Same as (ix), but without any exception. These last two criteria make use of
the expected correlated behaviour of the different indicators and at least (ix)
must be triggered for an automatic detection.

\begin{figure}
\begin{center}
\includegraphics[width=0.5\textwidth, clip]{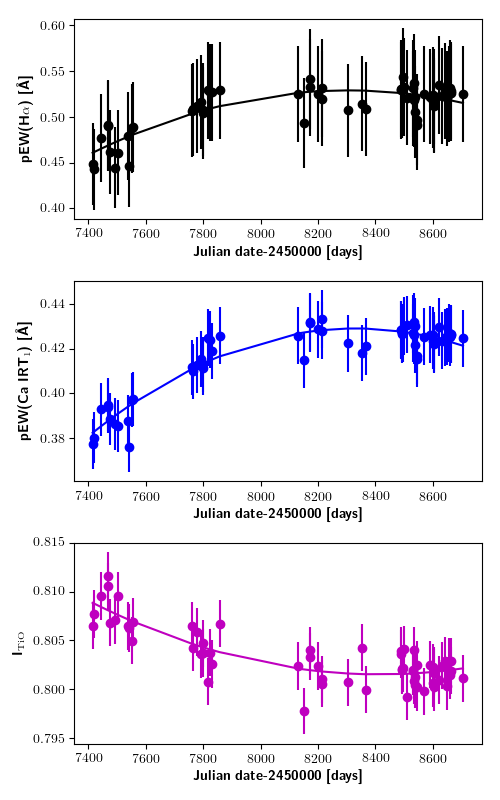}
        \caption{\label{fig:quad} Time series of J14257+236W / BD+24~2733A. 
        We show pEW(H$\alpha$) (\emph{top}) and
         pEW(\cairta) (\emph{bottom}). Additionally, the best-fit quadratic trend
         is indicated as the solid line.
}
\end{center}
\end{figure}

As an instructive example  of our quadratic trend search, we show the M0.0\,V 
star J14257+236W / BD+24~2733A in Fig. \ref{fig:quad}.
This clearly demonstrates the need of longer time baselines to clarify whether these quadratic trends lead to periodic behaviour or are just systematic episodes
embedded in chaotic variations.

\subsection{Searching for cycles}\label{sec:cycles}
To search for cyclic behaviour, we applied the generalised Lomb-Scargle (GLS) periodogram \citep{Zechmeister2009}
as implemented in 
PyAstronomy\footnote{\tt https://github.com/sczesla/PyAstronomy} \citep{pya} to each of our
chromospheric indicator time series, namely pEW(H$\alpha$), pEW(\cairta), pEW(\cairtb),
\itio, \rpr\, (whenever available), RV, CRX, and dLw.
In our further analysis, we searched for periods longer than  one year and shorter than 90\% of the time baseline of the individual data set 
to cover
whole cycles. We accepted periods with an FAP lower than 0.1\%
as significant. When we found significant periods in three or more indicators, we verified whether the
difference between the maximum peak periods of at least three indicators (not necessarily the 
significant ones) was less than 20\%, to ensure that the indicators showed (about) the same period.
In this case, we accepted the period as an automatic detection. We also included \rpr data in this procedure,
 but we caution that due to the usually much longer time baseline, we often found longer periods
 in these data as maximum peak. The search was therefore basically conducted on the CARMENES
 data alone.
Since we allow for a 20\% difference of the periods in the individual chromospheric indicators
we assign a 20\% error to all our measured cycles.

 In an extended search, we also accepted periods shorter than one year but longer than
  150\,days (to avoid rotation periods). This led to an automatic detection of nine additional
  stars, which were all rejected by visual inspection. They are usually caused by the observing
pattern or instead are weak quadratic trends. The only example of a star with a 0.7 a period, where the short period may be present
in the TiO index alone, is the M5.0 star J20260+585 / Wolf~1069, which we show in
the appendix in Fig. \ref{fig:M5}. Nevertheless, this GLS also shows a 
small peak in the window function that is caused by the observing scheme at about the period we found. We therefore also rejected this star by visual inspection.

All automatically found cycle candidates were inspected visually. This sometimes revealed that
the period we found can also be explained by a
quadratic or linear trend, especially when the period we found is near the length of the time
baseline of the observations. 

\begin{figure}
\begin{center}
\includegraphics[width=0.5\textwidth, clip]{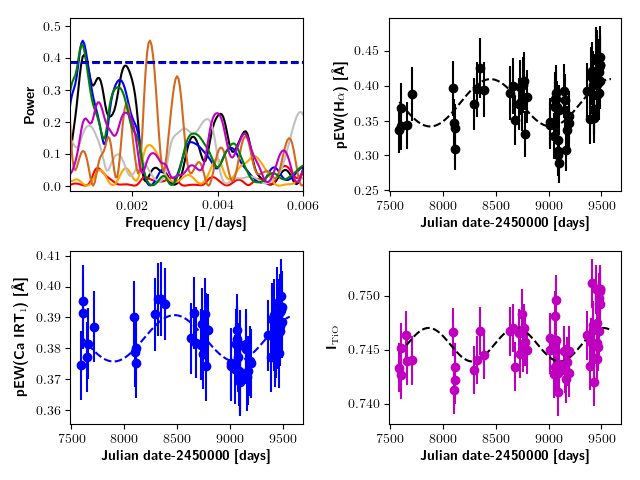}
        \caption{\label{fig:J22330} GLS and time series of J22330+093 / BD+08~4887. We 
         show the GLS (\emph{left, top}) with pEW(H$\alpha$) in black,
         pEW(\cairta) in blue, pEW(\cairtb) in green, the TiO index in magenta,
         RV in yellow, dLw in grey, CRX in orange, and the window function in red.
         The dashed blue line indicates FAP=0.01 for the pEW(\cairta) data.
         Additionally, we show the time series for pEW(H$\alpha$) (\emph{right, top}),
         for pEW(\cairta) (\emph{left, bottom}), and the TiO index (\emph{right, bottom}).
         The dashed black lines indicate the best sine fit for each indicator.
}
\end{center}
\end{figure}

An example of an formerly unknown cycle candidate that fulfilled our automatic detection
criteria and was not excluded by visual inspection is  the M0.1\,V star J22330+093/BD+08~4887,
for which we show the GLS and time series
in Fig. \ref{fig:J22330}. While pEW(H$\alpha$) and pEW(\cairt) do show about the same
period, the TiO index does not show the cycle period,
but a shorter period is preferred. Since the broad peak in the GLS reveals
that the period is not well constrained
and no \rpr\ data are available, the activity cycle
period needs to be confirmed by photometry or other spectroscopic data.

\section{Results and discussion}\label{sec:results}

With the methods described in the previous section, we found 26 automatic detections of
long-term variability in our sample
of 211 stars. Seven of these are linear trends, 14 are quadratic trends, and 12 are cycles  (some stars exhibit more than
one type of variability). From these numbers, we already excluded
one linear trend, one quadratic trend,
and one cycle, whose time series showed data gaps longer than 550 days and longer than 40\% of
  the observation time baseline. 
Our results are summarised in Table~\ref{tab:cycle},
where we list the Karmn number of the star,  a common name, the spectral type, the
number of spectra we used, and the time baseline of the CARMENES observations. Furthermore, in the case of automatically found linear trends, we list the highest Pearson correlation coefficient
(which does not necessarily correspond to the lowest or even a significant p-value),
and in case of a quadratic trend, the number of flags that were triggered. In the case of
significant GLS periods, we list the period length, and when available, the period length
found in the \rpr data. Additionally, we cite known literature values.

After the automatic detection, we inspected all variable time series by eye
to determine the most appropriate detection in cases of multiple detections.  For example,
in all three cases when we rejected the linear trend, a quadratic trend was  also found (twice by automatic detection, once by eye), and the quadratic trend was always preferred 
by visual evaluation. Six of the seven rejected cycles  also showed 
a quadratic trend (four by automatic detection, two by visual inspection). Two of the
stars with automatically detected quadratic trends (J04290+219 / BD+21~652, J05314-036 / HD~36395) also have a longer cycle detected in the \rpr\
data, and the cycle length we found is about the length of the observation time baseline.

Finally, four linear trends, 11 quadratic trends, and five cycle periods passed the
visual inspection. Therefore, 20 of our 26 automatically found long-term
  variable stars pass the visual inspection, which additionally leads to an appropriate
  categorisation. Since our adopted errors are relatively large, we refrained from 
  applying
  an F-test and performed the sorting by eye.
The stars that passed visual inspection are marked with an asterisk in Table~\ref{tab:cycle}, where we also
list some remarks from the visual inspection for many stars. 

\begin{sidewaystable*}
\caption{\label{tab:cycle} Proposed linear and quadratic trends and cycle periods for the stellar sample. }
\footnotesize
\begin{tabular}[h!]{llccccccccccc}
\hline
\hline
\noalign{\smallskip}

        Karmn           & Name &SpT$^{a}$   & No.    & Time &     Trend$^{b}$  & Para- & Cycle & \rpr $^{d}$ & Visual & \multicolumn{2}{c}{Literature}  \\
                &  &     & of  & baseline & corr-  & bolic$^{c}$     & auto  & cycle& inspection$^{e}$& \multicolumn{2}{c}{cycle}\\
                &        &     & spec.         & [a]   & coeff  &    & [a] & [a]  & & [a] & [a]\\
\noalign{\smallskip}
\hline
\noalign{\smallskip}
J09144+526& HD 79211& M0.0 &  153 &   2.7 &    0.693  &  . &  1.9 &  exist (13.2)&  nc, pa (Ca+H$\alpha$) &      &\\
J12123+544S& HD 238090& M0.0 &  104 &   3.3 &   ...     &  9 &  1.5 &  ...  & nc, pa*   &       & \\
J13450+176& BD+18 2776& M0.0 &   29 &   6.0 &   $-$0.653  & . & ... & ... & * \\
J14257+236W& BD+24 2733A& M0.0 &   63 &   3.5 &    0.829  &  10 &  1.2 &  ...  & nc, pa*      &  & \\
J02222+478& BD+47 612& M0.5 &   55 &   2.0 &   ...     &  9 &  ... &  exist$^{g}$ (14.0)&   *  &     lin trend Rob13\\
J04290+219& BD+21 652& M0.5 &  167 &   2.1 &   ...     &  10 &  1.8 &  3.4 (13.9)  & nc, pa*   & \\
J06105$-$218& HD 42581 A& M0.5 &   49 &   3.9 &   ...     &  9 &  2.7 &  14.3$^{f}$ (16.2) & cy*    &     8.4$\pm$0.3 SM16 &   8.3 DA19\\
J12312+086& BD+09 2636& M0.5 &   47 &   3.4 &   ...     &  9 &  ... &  ...  &    *  &     & \\
J13299+102& BD+11 2576& M0.5 &  354 &   6.1 &   ...     &  . &  1.5 &  0.7 (15.7) &  nc, pa (Ca)  &    & \\
J05415+534& HD 233153& M1.0 &   95 &   4.1 &    0.705  &  . &  1.4 &  ...  &   lin*+cy*      & \\
J16581+257& BD+25 3173& M1.0 &   54 &   1.1 &   ...     & 9 & ... & ... & *\\
J18051$-$030& HD 165222& M1.0 &   51 &   2.5 &   ...     &  9 &  ... &  7.6 (12.3) &       &         & \\
J22330+093 &BD+08 4887& M1.0 &   64  &  5.2 &   ...     &  .&   3.1 &    ...&  *\\
J23245+578&BD+57 2735&  M1.0 &   61 &   4.6 &   $-$0.740  &  . &  ... &    ...&  *   &\\
J02123+035& BD+02 348& M1.5 &   65 &   4.3 &   ...     &  9 &  ... &  15.5$^{f}$ (16.0)&   *         &\\
J03181+382& HD 275122& M1.5 &   56 &   4.9 &   $-$0.770  &  10 &  ... &  ...  &  pa*        &\\ 
J05314$-$036 & HD 36395& M1.5 &   90 &   2.2 &  ...      &  9 &  1.9 &  11.6$^{f}$ (16.3) &   nc, pa*    &     10.8/ 3.9 SM16&    3.5 K19\\
J16254+543& GJ 625& M1.5 &   30 &   1.3 &   ...     & 9 & ... & ...&* & 3.3$\pm$1.5 SM18\\
J22057+656& G 264-018 A& M1.5 &   90 &   2.6 &   ...     &  10 &  ... &  ...  &  *      &          & \\
J22565+165&HD 216899&  M1.5 &  548 &   6.1 &   ...     &  .&   5.4 &  15.5$^{f}$ (16.2)&    &      15.5 Per21     & \\
J11110+304W& HD 97101 B& M2.0 &   51 &   4.1 &   ...     &  9 &  ... & 10.1 (13.2) &  *   &       & \\
J01518+644& G 244-037& M2.5 &   21 &   5.2 &    0.705  &  . &  ... &  ...  &        *    &\\
J19169+051N& V1428 Aql& M2.5 &  125 &   1.7 &   ...     &  10 &  ... &  12.6$^{f}$ (14.3) &  &     3.3$\pm$0.4 DA19  &     9.3$\pm$1.9 SM16\\
J16167+672N& EW Dra & M3.0 &  103 &   4.5 &   ...     &  9 &  ... &  ...  &  *    &        &\\
J07274+052& Luyten's Star& M3.5 &  734 &   6.1 &   ...   &  . &  2.8 &  2.8$^{f}$ (14.8)  & *  & 6.6$\pm$1.3 SM16 &  2.8 Per21\\
J18346+401& LP 229-017& M3.5 &  77  &  4.8  &  ...      &  .  & 4.0  & ...   &  *     &          & \\
\noalign{\smallskip}                                                 
\hline                                             
                                                   
\end{tabular}                                      
\tablebib{                                         
  AK17:~\citet{AK17}; DA19:~\citet{DA19};    
  Per21:~\citet{Perdelwitz2021}; Rob13:~\citet{Robertson2013}; SM16:~\citet{Suarez-Mascareno2016}; SM18:~\citet{SuarezMascareno2018}. } \\
$^{a}$ Spectral (sub-)type of M \\
$^{b}$ Pearson's $r$ \\
$^{c}$ Number of flags triggered in the parabolic trend search\\
$^{d}$ Given in parentheses is the observation time baseline for the \rpr\ measurements\\
$^{e}$ Abbreviations: nc: no cycle, pa: parabolic trend, lin: linear trend,
*:Confirmed by visual inspection.\\
$^{f}$ Automatic detection of the period based on \rpr\ data alone, see Sect. \ref{sec:rpr}.\\
$^{g}$ exist: \rpr\, data exist, but the criterion of FAP$<0.001$ in the GLS is not met.
                               
\normalsize                    
\end{sidewaystable*}

\subsection{Comparison to literature}

For seven of the stars showing systematic variability, we could find published cycle
lengths. Unfortunately, the situation is complicated and we find only few agreements.
For J02222+478 / BD+47~612, we find a quadratic trend that agrees with the linear trend found by \citet{Robertson2013}.
 The same is true for J16254+543 / GJ~625, where we find a quadratic trend that agrees
with the cycle period found by \citet{SuarezMascareno2018}. For J05314-036 / HD~36395, a long and a short cycle is known. Since the time baseline of the CARMENES data is
about 1.5\,a shorter than the shorter cycle, the quadratic trend agrees with this
short cycle, while the period found in the \rpr\ data agrees with the long cycle.

For J19169+051N / V~1428 Aql, a short and a long cycle are known from photometry.
As the observation time baseline of the CARMENES data is rather short for this star, we
find a quadratic trend, while in the \rpr\ data, we find a period that roughly agees with the longer literature
period. For J22565+165 / HD~216899, we find a period of 5.4\,a without visual confirmation that is much shorter than the period found in the \rpr\ data.
 J06105-218 / HD~42581~A and J07274+052 / Luyten's star are discussed in Sect. \ref{sec:promising} because we
found a visually convincing cycle period.

\subsection{Promising activity cycle candidates}\label{sec:promising}

In the following, we discuss the five starst that are  the most promising candidates for activity cycles for which more than
a full cycle is covered by the CARMENES observations and that passed the visual inspection.
 The first of these stars is J22330+093 / BD+08~4887. This star was discussed in Sect. \ref{sec:cycles}, see Fig.~\ref{fig:J22330}.

\begin{figure*}
\begin{center}
\includegraphics[width=0.5\textwidth, clip]{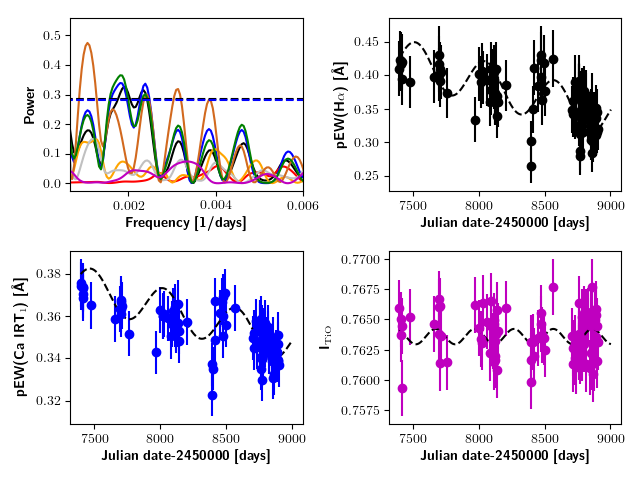}
\includegraphics[width=0.5\textwidth, clip]{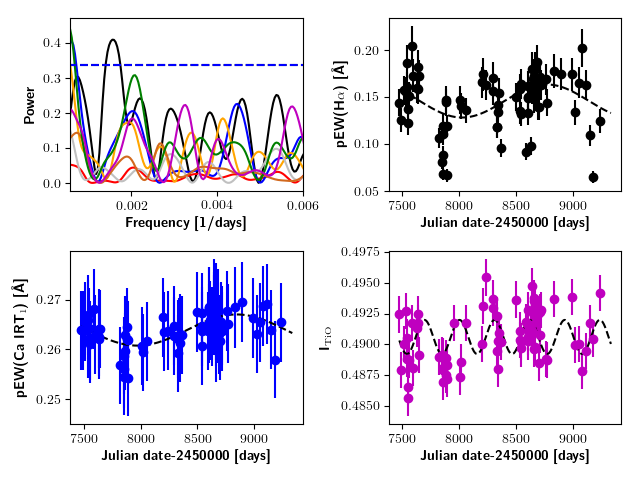}\\
\caption{\label{fig:cycle1} Same as in Fig.~\ref{fig:J22330}.
         GLS and time series of J05415+534 / HD~233153  (\emph{left}) and J18346+401 / LP~229-017 
         (\emph{right}). We note that for J05415+534 / HD~233153, a linear trend
           was found as well, which we include in the sine fit for pEW(H$\alpha$) and pEW(\cairta). We
           further note that for J18346+401 / LP~229-017, the highest peak of pEW(H$\alpha$) in
           the GLS is at about twice the frequency as for the other chromospheric indicators
           that triggered the cycle detection. We therefore used half the frequency for the sine fit
         of pEW(H$\alpha$) as well. 
}
\end{center}
\end{figure*}

In the
M1.0 dwarf J05415+534 / HD~233153 a rather short period of only 1.4\,a was found that we show in the left panel of Fig. \ref{fig:cycle1} in H$\alpha$ and \cairta. The trend that was also detected automatically is overlaid. It may indicate two different cycle periods.
However, the TiO index does not show any variation, not even the trend. 

For the M3.5\,V star J18346+401 / LP~229-017, we show the GLS and time series in the right panel of Fig. \ref{fig:cycle1}.
Although the period has about the same length as the observation time baseline, we opted for the cycle and not for a quadratic trend by visial inspection in this
special case. The period
can be seen quite clearly in H$\alpha$ and \cairt, but not in TiO, where a shorter
period is found.

 The case for the M0.5 dwarf J06105-218 / HD~42581~A is slightly more complicated. A cycle of about 8.3\,a
is known from the literature \citep{Suarez-Mascareno2016, DA19}. In the \rpr\ data, we find
a peak at about twice this period, which may be a sub-harmonic because there is also a (non-significant)
peak at 7.9\,a, which would agree with the literature value. In the CARMENES data, we also
find a cycle of 2.7\,a  length, which is visually confirmed. Either this is a quasi-periodic episode,
or this is a second, shorter cycle. We show the data of this star in Fig. \ref{fig:J06105}.

For the M3.5 dwarf J07274+052 / Luyten's star, we find a period of 2.8\,a, in agreement with the
findings from the \rpr\ data, which has been published previously \citep{Perdelwitz2021}. There is a second
published period of 6.6$\pm$1.3\,a \citep{Suarez-Mascareno2016} that may be about twice the shorter cycle.
A second significant peak lies at 6.1\,a in the GLS of the \rpr\ data. The correct period cannot be decided
from the available data, which are shown in Fig. \ref{fig:J07274}.


\subsection{Known cycles without detected long-term variation in the chromospheric indicators}

 We fail to find  systematic long-term activity variations with our automatic search algorithms for 34 stars with proposed cycles from the literature. These non-detections fall into
three categories: First, the star was excluded from our data analysis due to
binarity (1 star) or fewer than 20 observations (4 stars); or it showed possibly problematic
long data gaps  (1 stars).
Second, visual inspections revealed indications of linear or quadratic trends or even cycles
(18 stars) in individual chromospheric indicators. Third, we do not find any
systematic long-term variations (10 stars).
We list all these stars in Table \ref{tab:onlylit} and also remark on the visual inspection of the
CARMENES data.

An example that demonstrates the difficulties well is the
star J06371+175 / HD260655. We were able
to visually identify a linear trend in the pEW(\cairt) time series, while we find
a quadratic trend in pEW(H$\alpha$) and \itio. Since the baseline of the
CARMENES observations is 6.1\,a,
this neither agrees with the 7.3\,a cycle from the \rpr\, data nor with literature
values of 2-3\,a from \citet{DA19} or 4.9\,a from \citet{Suarez-Mascareno2016}.
 Since the trend we found in the data indicates a cycle that is longer than 10 years, the star may
exhibit multiple cycles. We caution, however, that the two cycle length values from the
literature (both photometry) also disagree, and claims of more than two cycles per star would
remain highly speculative given the available data. Alternatively, this may indicate quasi-periodic episodes or changes
in the cycle length, such as those known from the Sun.

Another instructive example is the star J06548+332 / Wolf~294/GJ~251. \citet{SN20} found a significant 600\,d cycle in RV data using HIRES
data alone, while the signal is not present in their CARMENES RV data. Their inspection 
of chromospheric indicators reveals a 660\,d cycle in the H$\alpha$ indicator and
a 300\,d cycle length 
in CRX and \ion{Na}{i}\,D.
While we find  a significant peak at about 630\,d in the TiO index and pEW(\cairtb), the GLS of the
pEW(H$\alpha$) shows an even
higher peak at about 1680\,d (including newer data),
while the \rpr\ data favour  about twice that period with a 3300\,d period, 
which may indicate that the period in pEW(H$\alpha$) is a harmonic, the longer period
of which cannot be found because the observation time baseline is too short. Further investigation is required to determine whether there 
 are indeed a short (600\,d=1.6\,a) and a long cycle (9\,a).

The star J11033+359 / Lalande~21185 is a peculiar case,
for which we find a visually convincing period
in
H$\alpha,$ but not in any other chromospheric indicator. This period was also
reported with a similar length by \citet{SN20} for their (shorter time baseline)
CARMENES data
alone. However, they replaced that period for a period that was more than twice longer
based on their combined CARMENES and SOPHIE data analysis. This value in turn ais bout
half the value we found from the \rpr\ data.

 On the other hand, there are examples such as J07446+035 / YZ~CMi, J10196+198 / AD~Leo, 
J17303+055 / BD+05~3409,  J20525$-$169 / LP816-060, or J22096$-$046 / BD$-$05~5715,
where we find a linear or quadratic
trend that agrees well with the  cycles
proposed in literature by visual inspection in single activity indicators. Moreover, the star J18580+059 /BD+05~3993 was discussed in
detail by \citet{SuarezMascareno2018}. Our observation started just after the time 
baseline discussed by them, but a downward trend was expected to occur afterwards, 
which is what we observe, but only by visual inspection.

Although the existence of multiple cycles in a single star may  explain
some of the discrepancies we found, the star may also undergo phases of
quasi-systematic episodes and therefore show a different behaviour at different times.
Another problem is that different activity indicators reveal different behaviour.
Trends or periods are often only found in one activity indicator, while the
others show a constant, chaotic, or even contradicting systematic behaviour. While the first two possibilities can be explained by the different sensitivity of the
indicators, the latter is not expected. It may indicate time lags between the
indicators in the (quasi-)systematic development. Possible examples for different
sensitivity are the stars J08161+013 / GJ~2066 and J11033+359 / Lalande~21185, for which we find
a quadratic trend or even a cycle in a single activity indicator, while the others are
rather constant. An example in which activity indicators show a different behaviour is 
J06548+332/Wolf~294 (discussed above, linear trend in pEW(\cairt), quadratic trend in pEW(H$\alpha$)). Since the discrepancy is seen only in the last observing block, this indicates a time lag between
the indicators.

 Another problem may be imposed by the observation scheme. An example is J08298+297 / DX~Cnc,
for which the observation pattern shows some tight blocks with larger gaps in between. Together 
with the rather high activity level of J08298+297 / DX~Cnc, this may be the cause for our finding only insignificant peaks in the GLS at about the known cycle period in pEW(H$\alpha$)
and pEW(\cairt).

\begin{sidewaystable*}
\caption{\label{tab:onlylit} Cycles from the literature where no long term variation was found by the
        automatic search. }
\footnotesize
\begin{tabular}[h!]{llcccccccccc}
\hline
\hline
\noalign{\smallskip}

        Karmn     & Name      & SpT$^{a}$   & No. of   & Time &     \rpr &  Visual & \multicolumn{3}{c}{Literature}\\
        &       &       & spectra  & baseline &    cycle    &   inspection   & \multicolumn{3}{c}{cycle}\\
         &                &      &          & [a]   & [a] & & [a] & [a] & [a]\\
\noalign{\smallskip}
\hline
\noalign{\smallskip}
J06371+175&  HD 260655& M0.0 &  115 &   6.1 &  7.3$^{c}$ (13.7)&      lin(Ca), pa(H$\alpha$,TiO);\rpr\ 2nd peak at 14.5\,a     & 2-3 DA19 & 4.9$\pm$0.4 SM16\\
J19346+045& BD+04 4157& M0.0&    53&    3.2&   ...  &                    &      3.0$\pm$0.8 DA19  &\\
J17303+055& BD+05 3409& M0.0&    53&    2.5&   11.9$^{c}$ (12.3)&    pa (H$\alpha$+Ca)   &      11.9 Per21    &\\
J18353+457& BD+45 2743&  M0.5&  16 &    1.2&   ... & too few data & 2.3$\pm$0.5 SM18\\
J18580+059& BD+05 3993& M0.5&    32 &   1.3&   exist (10.8) &  lin(H$\alpha$+Ca) & 5.6$\pm$0.1 SM18\\
J20451$-$313& AU Mic & M0.5&    90&    1.3&   ...  &            &      4.8,2.8 K19   &     40.6 BK20\\
J22021+014& BD+00 4810& M0.5 &  75  &  3.1  &   exist (14.3) & lin  & 8.5$\pm$1.1 SM16  & \\
J00051+457 & GJ 2&M1.0 &   52 &   1.9 &   exist (14.0)  &        & 3.2$\pm$0.1 SM18\\
J00183+440 & GX And&M1.0 &  213 &   4.1 &  exist (14.1)   & weak pa(H$\alpha$, TiO) & 2.8$\pm$0.5 SM18\\
J07361$-$031 & BD-02 2198 &M1.0 &   45 &   6.0 &  ...   &   known binary (Ba21)&         11.5$\pm$1.9 DA19\\
J22559+178& StKM 1-2065& M1.0&    11&    1.4&     ...&    too few data    &      2-3 DA19      &  4.4$\pm$0.9 SM18\\
J10122$-$037& AN Sex & M1.5&    74&    2.2&   2.6 (16.2)  &                &      3.2 $\pm$0.4 DA19 &   13.6$\pm$1.7 SM16\\
J11033+359& Lalande 21185 & M1.5 &  376 &   4.9 &   13.2$^{c}$ (14.9)& 3.2a cycle (H$\alpha$) &     7.9 SN20      \\
J13457+148& HD 119850& M1.5 &  249 &   2.2 &   exist (17.5) &      &9.9$\pm$2.8 SM16 & \\
J15218+209& OT Ser & M1.5&    52&    2.8&   7.3$^{c}$ (9.3) &                    &      6.5$\pm$0.8 DA19  &\\
J04429+189 & HD 285968& M2.0 &   23 &   1.0 &  7.3$^{c}$ (13.8)   &   weak lin trend       &     5.9$\pm$0.7 SM16&    5.6 GdS12\\
J08161+013& GJ 2066& M2.0 &   70 &   5.0 & 7.4 (11.3)  &    pa(Ca)  & 4.1$\pm$0.7 DA19  &\\
J11000+228& Ross 104 & M2.5&    55&    2.2&    0.7 (8.1) &   lin (TiO)  & 5.3$\pm$0.6 SM16 &\\
J12350+098& GJ 476 & M2.5&     9&    4.8&   ...  &    too few data     &      2.9 Rob13     &\\
J06548+332 & Wolf 294 & M3.0 &  342 &   6.1 &  9.0 (14.0)  &   4.5\,a cycle (H$\alpha$)  &     1.6 SN20&\\
J10196+198& AD Leo & M3.0 &   44 &   1.9 &  ...  &   data gap; lin    & 27 AK17  & 7.6 Buc14\\
J15194$-$077& HO Lib & M3.0&    52&    3.5&   4.0$^{c}$ (10.7) &    lin  (TiO)    &      4.5 K19/Rob13  &   6.2$\pm$0.9 SM16 &     3.85 GdS12\\
J08409$-$234& LP 844-008& M3.5 &   28 &   4.2 &   exist (13.0)&  lin (TiO+Ca), pa (H$\alpha$)  &     5.2$\pm$0.3 SM16 &\\
J13427+332& Ross1015 & M3.5&    19&    4.0&   ...  &    too few data     &      quadr trend Rob13&\\
J16303$-$126 & V2306 Oph&M3.5 &   88  &  3.5 &  4.2$^{c}$ (11.2)&  pa(H$\alpha$)  &  3.9$\pm$1.0 DA19  &     4.4$\pm$0.2 SM16\\
J22096$-$046& BD-05 5715& M3.5&    59&    3.4&   exist (15.0) &    pa (H$\alpha$+TiO)  &      10.2$\pm$0.9 SM16& \\
J18498$-$238& V1216 Sgr& M3.5&    51&    1.0&   exist (13.3) &           &      4.9 K19        &     7.1$\pm$0.1/2.1$\pm$0.1 SM16 &    4.1/1.0 IB21\\
J11477+008& FI Vir & M4.0&    51&    3.0&  exist (14.2) &    pa (TiO)   &      4.5$\pm$2.0 DA19  &     4.1$\pm$0.3 SM16\\
J20525$-$169& LP 816-060& M4.0&    35&    4.4&   ...  &    pa (H$\alpha$+Ca)   &      10.6$\pm$1.7 SM16& \\
J22532$-$142& IL Aqr& M4.0 &   67 &   3.5 &   exist (14.8) &        & 4.5$\pm$0.7 DA19  & \\
J07446+035& YZ CMi & M4.5&    48&    2.0&   ...  &    weak lin trend (H$\alpha$+Ca)     &      10.6$\pm$1.7 SM16&\\
J08413+594& LP 090-018& M5.5&    51&    3.9&   ...  &                   &      14 LS20       &\\
J10564+070& CN Leo & M6.0&    66&    2.2&   ...  &                   &      8.9$\pm$0.2 SM16 &\\
J08298+267& DX Cnc & M6.5&    29&    4.8&   ...  &    lin(TiO), 2.5a cycle insignificant (H$\alpha$,Ca)   &      2.67 BK20     &\\

\noalign{\smallskip}
\hline                              
\end{tabular}     
\tablebib{        
  Ba21:~\citet{Baroch2021}; BK20:~\citet{BK20}; Buc14:~\citet{Buccino2014}; 
  DA19:~\citet{DA19};
  GdS12:~\citet{GdS12}; \\IB21:~\citet{IB21}; K19:~\citet{Kueker2019}; LS20:~\citet{Lopez-Santiago2020}
        Per21:~\citet{Perdelwitz2021}; Rob13:~\citet{Robertson2013}; \\ SM16:~\citet{Suarez-Mascareno2016}; SM18:~\citet{SuarezMascareno2018}; SN20:~\citet{SN20}.} 
\\
$^{a}$ Spectral (sub-)type of M \\
$^{b}$ Given in parentheses is the observation time baseline for the \rpr\ measurements\\
$^{c}$ Automatic detection of the period based on \rpr\ data alone, see Sect. \ref{sec:rpr}.\\

\normalsize

\end{sidewaystable*}

\subsection{Promising cycle candidates from \rpr\ alone}\label{sec:rpr}

Since the \rpr\ data usually show a much longer observation time baseline and \rpr\ is known to be suitable for activity cycle search in more solar-type stars
\citep{Baliunas1995, Brandenburg2017}, we decided to perform an additional search on the \rpr\
data alone. This can be achieved even though the data originate from seven different telescopes,
  because instrument offsets were minimised by extracting the \rpr\ values using a direct comparison to model spectra (for further discussion, see
  \citet{Perdelwitz2021}). Since only a single chromospheric indicator was used this time and a
  cross-check with other indicators is therefore missing,
  we opted in this search for a much better FAP $< 0.0005$ and excluded all stars
with fewer than 20 spectra or automatically detected periods shorter than one year. Since the \rpr\
data often have a much coarser sampling, we also excluded stars with data gaps longer than a quarter
of the observation time baseline and longer than 1000 days.

We repeated the GLS analysis and found that 22 cycle candidates fulfilled these criteria. Out of these, 7
correspond to stars that also show a variability detection in the CARMENES data and are therefore listed
in Table \ref{tab:cycle}. Another 6 correspond to stars with a proposed cycle from the literature
and are therefore listed in Table \ref{tab:onlylit}. A cycle period detection based on the \rpr\ data
alone is indicated in both of the tables. The remaining 9 cycle candidates are detections
based on \rpr\ alone, and none of these was rejected by visual inspection. We list them
in Table \ref{tab:volker} with their spectral type, number of observations, the \rpr\ observation time baseline,
and remarks on the visual inspection of the CARMENES data. 
We show the GLS and
time series of \rpr\ of the most promising 6  candidates in the appendix in Figs. \ref{fig:rpalone1}, \ref{fig:rpalone2},
and \ref{fig:rpalone3}.

 This better performance of the \rpr\ data in the cycle search is partly caused
  by the longer time baseline, which allows us to find cycles that are not yet covered by the CARMENES data.
  We therefore state the CARMENES baselines in Table \ref{tab:volker}, where appropriate. Nevertheless,
  in some of these cases, no trends are found, either. While the reason may be that the
  CARMENES observations incidentally cover the maximum or minimum of the cycle,
  the different sensitivity of the lines may also play a role. We discuss this in Sect. \ref{sec:sens}
  and reveal that the relative amplitude in \rpr\ data is higher than of the other line pEWs.

\begin{table*}
\caption{\label{tab:volker} Proposed cycle periods using \rpr\ alone for stars without CARMENES data detections or literature values. }
\footnotesize
\begin{tabular}[h!]{llccccccccccc}
\hline
\hline
\noalign{\smallskip}

        Karmn  &Name         & SpT$^{a}$   & No.    & Time &    \rpr & Visual   \\
        &       &      & of  & baseline &  cycle& inspection\\
        &                &     & spec.         & [a]   & [a] \\
\noalign{\smallskip}
\hline
\noalign{\smallskip}
J10251$-$102 &BD-09 3070& M1.0  &  62 &   9.0  &  2.3 & 6.1\,a CARMENES baseline, weak trend in Ca, H$\alpha$\\
J11054+435 &BD+44 2051A& M1.0  & 125 &   8.1  &  1.8 & data gap, 2.7\,a in Ca, H$\alpha$$^{b}$ \\
J14010$-$026 &HD 122303& M1.0  & 226 &   11.7 &  2.6  & 1.3\,a CARMENES baseline, no trend\\
J20533+621 &HD 199305& M1.0  & 73 &    14.1 & 7.8 & 3.8\,a CARMENES baseline, 3.2\,a in H$\alpha$$^{b}$\\
J23492+024 &BR Psc& M1.0  & 207 &   14.2 & 27.3 & 5.6\,a CARMENES baseline, maybe trend in H$\alpha$\\
J20450+444 &BD+44 3567& M1.5  & 80  &   6.7  & 7.2 & 3.7\,a in Ca, H$\alpha$$^{b}$\\
J10289+008 &BD+01 2447& M2.0  & 204 &   12.9 &  3.3 & 3.1\,a in Ca, H$\alpha$$^{b}$ \\
J14294+155 &Ross 130& M2.0  & 52  &   7.7  &  1.9 & too few CARMENES data\\
J11421+267 &Ross 905& M2.5  & 302 &   15.2 &  17.2 & 2.3\,a CARMENES baseline, no trend\\
                                                                                                              
\noalign{\smallskip}                                                 
\hline                                             
                                                   
\end{tabular} \\                                     
$^{a}$ Spectral (sub-)type of M \\
$^{b}$periods given correspond to the highest peak in the periodogram, regardless of the assigned FAP.                           
\normalsize                    
\end{table*}

\subsection{Dependence on spectral type}\label{sec:spectype}

\begin{figure}
\begin{center}
\includegraphics[width=0.5\textwidth, clip]{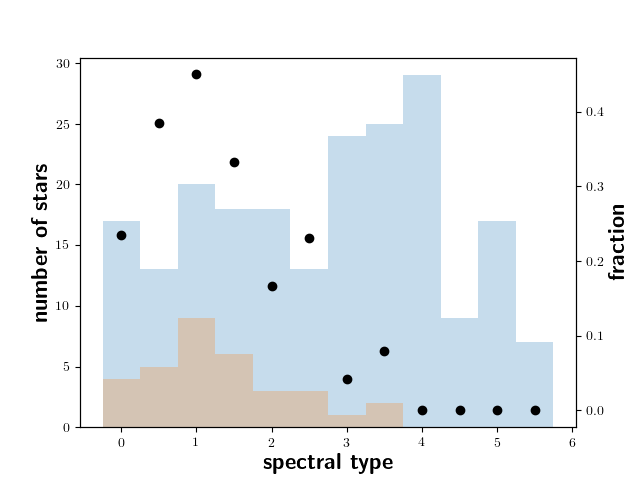}
        \caption{\label{fig:spectralbehaviour} Number of stars 
        per spectral sub-type
        (blue bars,  left y-axis), and number of stars with an automatic variability detection (red bars,  left y-axis).
         We show  the
fraction of stars with variability per spectral
        sub-type as black dots corresponding to the right y-axis.
}
\end{center}
\end{figure}

To identify a possible dependence on spectral sub-type, we show in Fig. 
\ref{fig:spectralbehaviour} the number of stars and the number of automatic long-term variability detections (including the nine detections based on \rpr\ alone).
The percentage of systematically variable stars 
drops toward later spectral types from about 15-30\% in early-M dwarf stars to lower than 5\%
for mid-M dwarfs.
We do not find any sign of a systematic long-term variability for any star
with spectral type later than  M4.0\,V. This cannot be caused by observational biases because the mean observation time baseline of the included M4 stars is even longer than for M0
  stars. Although the mean number of observations for M4 stars is somewhat lower than for M0
  stars, the number of stars with more than 50 usable observations is higher for M4 stars than
  for M0 stars. Noise should not play a role here either because we excluded spectra with a too low level of the signal-to-noise
 ratio, and this is usually only an issue for stars with spectral type M5 and later because the exposure time
  never exceeds 1800\,s.  Either these latest type, fully convective stars
do not have activity cycles, or they may not found due to frequent flaring.

Flaring was also identified as the main reason for variability in fully convective M dwarfs in a
study by \citet{Medina2022}, who found only for one very young star
a correlation of the EW(H$\alpha$) to the rotational phase  in
a sample of 13 stars. The short timescales of steallar variability in their study, which was observed with high cadence, suggest an
origin of the variability in flares. 
Although we excluded large flares, frequent low-level flaring would
lead to a broadening of the distribution of the individual indicators and
would thus not be handled by $\sigma$ clipping. This can  lead
to a veiling of possible activity cycles. 

The stars with cycle values proposed in the literature also include stars with spectral
  type M4 and later. We therefore note that all these stellar cycles have been detected by photometry
  and not by chromospheric indicators. At least for J07446+035 / YZ~CMi and J20525-169 / LP 816-060,
  visual inspection also revealed weak linear or parabolic trends in these stars,
  while for J08298+297 / DX~Cnc, we find an insignificant period of the cycle length from the literature. This may indicate that
  in these relatively late-M dwarfs, the sensitivity of the chromospheric indicators
  to cyclic activity is lower than in early-type M dwarfs. Next to flaring, this may be caused by generally high chromospheric filling factors, for example, which would lead to 
low amplitudes in the cycle modulation and might therefore hamper the detection of a cycle. 

\subsection{Sensitivity of the different indicators}\label{sec:sens}

We analysed the chromospheric indicator that was
most often involved in the trend or cycle detection for the different search methods. For the linear trend search, the
\cairt\ lines, H$\alpha,$ and dLw most often led to a trend with between 5 and 7 
detections each, while the TiO  index and CRX led to one detection,
and RV to zero detection. For the parabolic search, we find
that pEW(H$\alpha$), pEW(\cairt), and \itio\, each were triggered for all detected stars.
Finally, for the cycle search,
again the \cairt\ lines were involved most often in a cycle detection (12 detections each),
followed by H$\alpha$ with 8 detections, dLw with 6 detections,
TiO and CRX  with 4 detections,
and
RV with 3 detections.
We conclude that long-term variability is most easily found in the \cairt\ lines,
followed by H$\alpha$ and dLw, while TiO, RV, and CRX are less well suited.

For a better comparison, we also considered the amplitudes of the different indicators. We computed
the amplitude as the relative difference of the median of the ten highest and ten lowest data points in
the time series with a detected (linear or quadratic) trend or period (regardless of whether the
indicator triggered the detection). For the time series
showing a trend, the amplitude is only a lower threshold. This leads to median amplitudes
of 14\% for pEW(H$\alpha$), of 4\% for pEW(\cairt), and of 0.08\% for \itio. This underlines
the higher sensitivity of H$\alpha$ compared to \cairt\ and especially \itio. The highest
sensitivity is found for \rpr, with a relative median amplitude of 46\%.
The relatively low amplitudes of pEW(\cairt) come as a surprise because \cairt\ triggered the
most automatic detections. This better performance compared to pEW(H$\alpha$) may be caused
by a larger influence of flares on the pEW(H$\alpha$), because the line is formed at
larger heights in the chromosphere than \cairt.



\subsection{Comparison of cycles to rotation periods}
While \citet{BoehmVitense2007} found two distinct relations between rotation period
and cycle length for active and inactive G and K stars, \citet{Kueker2019} found
virtually no dependence of the cycle length on rotation period for early-M dwarfs.
They found two groups with characteristic cycle times, however, first the fast rotators
with rotation periods shorter than one day and cycle lengths of about one year, and 
second, stars with longer rotation periods and a cycle length between three to five years.
Moreover, \citet{Suarez-Mascareno2016} did not find a correlation between rotation period and cycle length
for the M dwarfs included in their study. Nevertheless, they find a weak correlation for
earlier-type stars.  \citet{SuarezMascareno2018} found a weak correlation
of rotation period with cycle length for M dwarfs using literature values in addition to 
their measurements, but also stated that a correlation test showed that there is a 37\% probability of the correlation being spurious.

Although we have 25 stars with known rotation periods shorter than one day in our sample,
these stars are all of spectral type M3.5 or later, and no long-term variation is
found for any of these stars. If the finding of \citet{Kueker2019} is correct that such short
cycles are found only for fast rotators, our visual evaluation
that none of the found periods shorter than one year is also consistent because none of these stars is a fast rotators.

 We searched for a correlation of rotation period $P_{\mathrm{rot}}$ and cycle length $
  P_{\mathrm{cyc}}$ using our
cycle detections from CARMENES and \rpr\ data and all literature values from
Table \ref{tab:onlylit}. This resulted in 34 stars (13 from our own measurements and 21 from the literature) for which rotation periods are also available.
We performed an analysis similar to that of \citet{SuarezMascareno2018} and computed the 
slope of $\log (P_{\mathrm{cyc}}/P_{\mathrm{rot}})$ versus 
$\log(1/P_{\mathrm{rot}})$ , which gives $1.1\pm0.1$ and therefore does not deviate significantly from 1.0. This implies that
there is no correlation between rotation period and cycle length.



\section{Conclusions}\label{conclusion}

We searched in a sample of 211 M dwarfs for long-term systematic variability
in chromospheric indicators. The observations were taken from the 
more than 19\,000 CARMENES
observations, considering only stars with more than 19 observations and an
observation time baseline of more than 200 days. We not only considered the pEW of H$\alpha$ and the \cairt\, lines as chromospheric indicators, but also more recently
defined activity indicators: the TiO-index, RV, CRX, and dLw. In a sub-sample
of 186 stars, we also used already published \rpr data with much longer observation
time baselines, originating from different instruments.
In these we performed an automatic search for periods and other long-term systematics,
namely  linear trends and quadratic trends. The automatic search
led to 26 detections of this systematic long-term variability. By visual inspection,
we rejected 6 of these detections and sorted the remaining stars uniquely into our different
variability categories, leaving us with four linear trends,
11 quadratic trends, and five  periods. We show all the
five stars with proposed periods in the Figs.~\ref{fig:J22330}, \ref{fig:cycle1}, 
\ref{fig:J06105}, and \ref{fig:J07274}. 
Analysing the \rpr\ data alone and applying a stricter FAP threshold than for
the combined data set, we found an
additional nine promising cycle candidates, which are shown in Figs. \ref{fig:rpalone1}, \ref{fig:rpalone2},
and \ref{fig:rpalone3}.
Furthermore, we found that from the analysed indicators that are available from the CARMENES observations,
the pEWs of \cairt\ and H$\alpha$ lines are best suited for a study of
long-term variability.  Comparison of the relative median amplitudes of the systematic variation
shows that \rpr\ has the highest amplitude, followed by pEW(H$\alpha$), pEW(\cairt), 
and \itio.  Therefore, we propose that future activity cycle searches performed in chromospheric
indicators should concentrate on \rpr, pEW(H$\alpha$), or pEW(\cairt). On the other
hand, we were unable to confirm many cycles proposed in the literature on the basis of photometric data, especially
for later-type stars, which may suggest that photometry is even better suited for an activity
cycle search than chromospheric indicators in these stars. Most important for RV exoplanet searches, we
found that
RV variations are not a sensitive indicator of systematic long-term chromospheric variability
in our study. 

 The detection rate of
systematic long-term variability in all our sample stars is about 12\% from the
CARMENES data alone.  This is low in comparison to \citet{SuarezMascareno2018},
for instance,  who found cycles in about 18\% of their M0--M3 stars using chromospheric indicators as well. When we only consider our M0--M3 stars,
  we have an even slightly higher detection rate of about 25\%, which is to be expected because
  we additionally searched for linear and quadratic trends. The detection rate of the long-term variability dramatically
decreases along the M spectral sequence to later-type stars. For stars M4.0 and later, we do 
not find any activity cycle. These fully convective stars may either exhibit no cycles 
caused by the different dynamo underlying their activity phenomena, or the existence 
of cycles is veiled in the chromospheric indicators by frequent flaring. This latter possibility is underlined by the study of \citet{Suarez-Mascareno2016}, who found cycles in 
about 50\% of their M dwarfs for fully convective stars as well, using photometry. Again, this suggests that photometry is better suited for cycle searches in M dwarfs.

Although the CARMENES observation span little more than six years at most, we found
some short-cycle candidates in agreement with literature values, or promising
parts of cycles where sometimes longer (mostly photometrically determined) cycles were
already known. We also identified a few cases for which modulation periods
reported in the literature could
not be confirmed. The most promising linear and quadratic trends  stress
that the observations need to be continued in the future to accumulate
data that cover at least a whole cycle or reveal the current data to be a quasi-systematic
episode.

\begin{acknowledgements}
  B.~F. acknowledges funding by the DFG under Schm \mbox{1032/69-1}.
  We thank our referee for the careful reading and the suggestions for improvement.
        CARMENES is an instrument for the Centro Astron\'omico Hispanoen Andaluc\'ia de
  Calar Alto (CAHA, Almer\'{\i}a, Spain). 
We acknowledge financial support from the Deutsche Forschungsgemeinschaft
(DFG) through the priority programme SPP 1992 ``Exploring the Diversity of
Extrasolar Planets'' (JE 701/5-1), and from the Agencia Estatal de
Investigaci\'on 10.13039/501100011033 of the Ministerio de Ciencia e
Innovaci\'on and the ERDF ``A way of making Europe'' through projects
PID2019-109522GB-
C5[1:4], PGC2018-098153-B-C33,
and the Centre of Excellence ``Severo Ochoa'' and ``Mar\'ia de Maeztu''
awards to the Instituto de Astrof\'isica de Canarias (CEX2019-000920-S),
Instituto de Astrof\'isica de Andaluc\'ia (SEV-2017-0709), and Centro de
Astrobiolog\'ia (MDM-2017-0737), and the Generalitat de Catalunya/CERCA
programme.
  CARMENES is funded by the German Max-Planck-Gesellschaft (MPG), 
  the Spanish Consejo Superior de Investigaciones Cient\'{\i}ficas (CSIC),
  the European Union through FEDER/ERF FICTS-2011-02 funds, 
  and the members of the CARMENES Consortium 
  (Max-Planck-Institut f\"ur Astronomie,
  Instituto de Astrof\'{\i}sica de Andaluc\'{\i}a,
  Landessternwarte K\"onigstuhl,
  Institut de Ci\`encies de l'Espai,
  Institut f\"ur Astrophysik G\"ottingen,
  Universidad Complutense de Madrid,
  Th\"uringer Landessternwarte Tautenburg,
  Instituto de Astrof\'{\i}sica de Canarias,
  Hamburger Sternwarte,
  Centro de Astrobiolog\'{\i}a and
  Centro Astron\'omico Hispano-Alem\'an), 
  with additional contributions by the Spanish Ministry of Economy, 
  the German Science Foundation through the Major Research Instrumentation 
    Programme and DFG Research Unit FOR2544 ``Blue Planets around Red Stars'', 
  the Klaus Tschira Stiftung, 
  the states of Baden-W\"urttemberg and Niedersachsen, 
  and by the Junta de Andaluc\'{\i}a.

\end{acknowledgements}

\bibliographystyle{aa}
\bibliography{papers}

\begin{thebibliography}{54}
\expandafter\ifx\csname natexlab\endcsname\relax\def\natexlab#1{#1}\fi

\bibitem[{{Alekseev} \& {Kozhevnikova}(2017)}]{AK17}
{Alekseev}, I.~Y. \& {Kozhevnikova}, A.~V. 2017, Astronomy Reports, 61, 221

\bibitem[{{Alonso-Floriano} {et~al.}(2015){Alonso-Floriano}, {Morales},
  {Caballero}, {Montes}, {Klutsch}, {Mundt}, {Cort{\'e}s-Contreras}, {Ribas},
  {Reiners}, {Amado}, {Quirrenbach}, \& {Jeffers}}]{AF15a}
{Alonso-Floriano}, F.~J., {Morales}, J.~C., {Caballero}, J.~A., {et~al.} 2015,
  \aap, 577, A128

\bibitem[{{Baliunas} {et~al.}(1995){Baliunas}, {Donahue}, {Soon}, {Horne},
  {Frazer}, {Woodard-Eklund}, {Bradford}, {Rao}, {Wilson}, {Zhang}, {Bennett},
  {Briggs}, {Carroll}, {Duncan}, {Figueroa}, {Lanning}, {Misch}, {Mueller},
  {Noyes}, {Poppe}, {Porter}, {Robinson}, {Russell}, {Shelton}, {Soyumer},
  {Vaughan}, \& {Whitney}}]{Baliunas1995}
{Baliunas}, S.~L., {Donahue}, R.~A., {Soon}, W.~H., {et~al.} 1995, \apj, 438,
  269

\bibitem[{{Ballester} {et~al.}(2002){Ballester}, {Oliver}, \&
  {Carbonell}}]{Ballester2002}
{Ballester}, J.~L., {Oliver}, R., \& {Carbonell}, M. 2002, \apj, 566, 505

\bibitem[{{Baroch} {et~al.}(2021){Baroch}, {Morales}, {Ribas}, {B{\'e}jar},
  {Reffert}, {Cardona Guill{\'e}n}, {Reiners}, {Caballero}, {Quirrenbach},
  {Amado}, {Anglada-Escud{\'e}}, {Colom{\'e}}, {Cort{\'e}s-Contreras},
  {Dreizler}, {Galad{\'\i}-Enr{\'\i}quez}, {Hatzes}, {Jeffers}, {Henning},
  {Herrero}, {Kaminski}, {K{\"u}rster}, {Lafarga}, {Lodieu},
  {L{\'o}pez-Gonz{\'a}lez}, {Montes}, {Pall{\'e}}, {Perger}, {Pollacco},
  {Rodr{\'\i}guez-L{\'o}pez}, {Rodr{\'\i}guez}, {Rosich}, {Sch{\"o}fer},
  {Schweitzer}, {Shan}, {Tal-Or}, \& {Zechmeister}}]{Baroch2021}
{Baroch}, D., {Morales}, J.~C., {Ribas}, I., {et~al.} 2021, \aap, 653, A49

\bibitem[{{Baroch} {et~al.}(2018){Baroch}, {Morales}, {Ribas}, {Tal-Or},
  {Zechmeister}, {Reiners}, {Caballero}, {Quirrenbach}, {Amado}, {Dreizler},
  {Lalitha}, {Jeffers}, {Lafarga}, {B{\'e}jar}, {Colom{\'e}},
  {Cort{\'e}s-Contreras}, {D{\'{\i}}ez-Alonso},
  {Galad{\'{\i}}-Enr{\'{\i}}quez}, {Guenther}, {Hagen}, {Henning}, {Herrero},
  {K{\"u}rster}, {Montes}, {Nagel}, {Passegger}, {Perger}, {Rosich},
  {Schweitzer}, \& {Seifert}}]{Baroch2018}
{Baroch}, D., {Morales}, J.~C., {Ribas}, I., {et~al.} 2018, \aap, 619, A32

\bibitem[{{B{\"o}hm-Vitense}(2007)}]{BoehmVitense2007}
{B{\"o}hm-Vitense}, E. 2007, \apj, 657, 486

\bibitem[{{Bondar'} \& {Katsova}(2020)}]{BK20}
{Bondar'}, N.~I. \& {Katsova}, M.~M. 2020, Geomagnetism and Aeronomy, 60, 942

\bibitem[{{Boro Saikia} {et~al.}(2018){Boro Saikia}, {Marvin}, {Jeffers},
  {Reiners}, {Cameron}, {Marsden}, {Petit}, {Warnecke}, \&
  {Yadav}}]{BoroSaikia2018}
{Boro Saikia}, S., {Marvin}, C.~J., {Jeffers}, S.~V., {et~al.} 2018, \aap, 616,
  A108

\bibitem[{{Brandenburg} {et~al.}(2017){Brandenburg}, {Mathur}, \&
  {Metcalfe}}]{Brandenburg2017}
{Brandenburg}, A., {Mathur}, S., \& {Metcalfe}, T.~S. 2017, \apj, 845, 79

\bibitem[{{Brown} {et~al.}(2022){Brown}, {Jeffers}, {Marsden}, {Morin}, {Boro
  Saikia}, {Petit}, {Jardine}, {See}, {Vidotto}, {Mengel}, {Dahlkemper}, \&
  {the BCool Collaboration}}]{Brown2022}
{Brown}, E.~L., {Jeffers}, S.~V., {Marsden}, S.~C., {et~al.} 2022, \mnras, 514,
  4300

\bibitem[{{Buccino} {et~al.}(2014){Buccino}, {Petrucci}, {Jofr{\'e}}, \&
  {Mauas}}]{Buccino2014}
{Buccino}, A.~P., {Petrucci}, R., {Jofr{\'e}}, E., \& {Mauas}, P. J.~D. 2014,
  \apjl, 781, L9

\bibitem[{{Caballero} {et~al.}(2016){Caballero}, {Gu{\`a}rdia}, {L{\'o}pez del
  Fresno}, {Zechmeister}, {de Juan}, {Alonso-Floriano}, {Amado}, {Colom{\'e}},
  {Cort{\'e}s-Contreras}, {Garc{\'{\i}}a-Piquer}, {Gesa}, {de Guindos},
  {Hagen}, {Helmling}, {Hern{\'a}ndez Casta{\~n}o}, {K{\"u}rster},
  {L{\'o}pez-Santiago}, {Montes}, {Morales Mu{\~n}oz}, {Pavlov}, {Quirrenbach},
  {Reiners}, {Ribas}, {Seifert}, \& {Solano}}]{Caballero2}
{Caballero}, J.~A., {Gu{\`a}rdia}, J., {L{\'o}pez del Fresno}, M., {et~al.}
  2016, in \procspie, Vol. 9910, Observatory Operations: Strategies, Processes,
  and Systems VI, 99100E

\bibitem[{{Coffaro} {et~al.}(2020){Coffaro}, {Stelzer}, {Orlando}, {Hall},
  {Metcalfe}, {Wolter}, {Mittag}, {Sanz-Forcada}, {Schneider}, \&
  {Ducci}}]{Coffaro2020}
{Coffaro}, M., {Stelzer}, B., {Orlando}, S., {et~al.} 2020, \aap, 636, A49

\bibitem[{{Cram} \& {Mullan}(1979)}]{Cram1979}
{Cram}, L.~E. \& {Mullan}, D.~J. 1979, \apj, 234, 579

\bibitem[{{Czesla} {et~al.}(2019){Czesla}, {Schr{\"o}ter}, {Schneider},
  {Huber}, {Pfeifer}, {Andreasen}, \& {Zechmeister}}]{pya}
{Czesla}, S., {Schr{\"o}ter}, S., {Schneider}, C.~P., {et~al.} 2019, {PyA:
  Python astronomy-related packages}, Astrophysics Source Code Library, record
  ascl:1906.010

\bibitem[{{D{\'\i}ez Alonso} {et~al.}(2019){D{\'\i}ez Alonso}, {Caballero},
  {Montes}, {de Cos Juez}, {Dreizler}, {Dubois}, {Jeffers}, {Lalitha}, {Naves},
  {Reiners}, {Ribas}, {Vanaverbeke}, {Amado}, {B{\'e}jar},
  {Cort{\'e}s-Contreras}, {Herrero}, {Hidalgo}, {K{\"u}rster}, {Logie},
  {Quirrenbach}, {Rau}, {Seifert}, {Sch{\"o}fer}, \& {Tal-Or}}]{DA19}
{D{\'\i}ez Alonso}, E., {Caballero}, J.~A., {Montes}, D., {et~al.} 2019, \aap,
  621, A126

\bibitem[{{Dumusque} {et~al.}(2011){Dumusque}, {Udry}, {Lovis}, {Santos}, \&
  {Monteiro}}]{Dumusque2011}
{Dumusque}, X., {Udry}, S., {Lovis}, C., {Santos}, N.~C., \& {Monteiro},
  M.~J.~P.~F.~G. 2011, \aap, 525, A140

\bibitem[{{Gleissberg}(1945)}]{Gleissberg1945}
{Gleissberg}, W. 1945, The Observatory, 66, 123

\bibitem[{{Gomes da Silva} {et~al.}(2012){Gomes da Silva}, {Santos}, {Bonfils},
  {Delfosse}, {Forveille}, {Udry}, {Dumusque}, \& {Lovis}}]{GdS12}
{Gomes da Silva}, J., {Santos}, N.~C., {Bonfils}, X., {et~al.} 2012, \aap, 541,
  A9

\bibitem[{{Hauschildt} {et~al.}(1999){Hauschildt}, {Allard}, \&
  {Baron}}]{Hauschildt1999}
{Hauschildt}, P.~H., {Allard}, F., \& {Baron}, E. 1999, \apj, 512, 377

\bibitem[{{Husser} {et~al.}(2013){Husser}, {Wende-von Berg}, {Dreizler},
  {Homeier}, {Reiners}, {Barman}, \& {Hauschildt}}]{Husser2013}
{Husser}, T.-O., {Wende-von Berg}, S., {Dreizler}, S., {et~al.} 2013, \aap,
  553, A6

\bibitem[{{Iba{\~n}ez Bustos} {et~al.}(2021){Iba{\~n}ez Bustos}, {Buccino}, \&
  {Mauas}}]{IB21}
{Iba{\~n}ez Bustos}, R.~V., {Buccino}, A.~P., \& {Mauas}, P.~J.~D. 2021, in The
  20.5th Cambridge Workshop on Cool Stars, Stellar Systems, and the Sun
  (CS20.5), Cambridge Workshop on Cool Stars, Stellar Systems, and the Sun, 47

\bibitem[{{Jeffers} {et~al.}(2022{\natexlab{a}}){Jeffers}, {Barnes},
  {Sch{\"o}fer}, {Quirrenbach}, {Zechmeister}, {Amado}, {Caballero},
  {Fern{\'a}ndez}, {Rodr{\'\i}guez}, {Ribas}, {Reiners}, {Cardona Guill{\'e}n},
  {Cifuentes}, {Czesla}, {Hatzes}, {K{\"u}rster}, {Montes}, {Morales},
  {Pedraz}, \& {Sadegi}}]{Jeffers2022a}
{Jeffers}, S.~V., {Barnes}, J.~R., {Sch{\"o}fer}, P., {et~al.}
  2022{\natexlab{a}}, \aap, 663, A27

\bibitem[{{Jeffers} {et~al.}(2022{\natexlab{b}}){Jeffers}, {Cameron},
  {Marsden}, {Boro Saikia}, {Folsom}, {Jardine}, {Morin}, {Petit}, {See},
  {Vidotto}, {Wolter}, \& {Mittag}}]{Jeffers2022}
{Jeffers}, S.~V., {Cameron}, R.~H., {Marsden}, S.~C., {et~al.}
  2022{\natexlab{b}}, \aap, 661, A152

\bibitem[{{Jeffers} {et~al.}(2018){Jeffers}, {Mengel}, {Moutou}, {Marsden},
  {Barnes}, {Jardine}, {Petit}, {Schmitt}, {See}, {Vidotto}, \& {BCool
  Collaboration}}]{Jeffers2018a}
{Jeffers}, S.~V., {Mengel}, M., {Moutou}, C., {et~al.} 2018, \mnras, 479, 5266

\bibitem[{{K{\"u}ker} {et~al.}(2019){K{\"u}ker}, {R{\"u}diger}, {Olah}, \&
  {Strassmeier}}]{Kueker2019}
{K{\"u}ker}, M., {R{\"u}diger}, G., {Olah}, K., \& {Strassmeier}, K.~G. 2019,
  \aap, 622, A40

\bibitem[{{Lafarga} {et~al.}(2021){Lafarga}, {Ribas}, {Reiners}, {Quirrenbach},
  {Amado}, {Caballero}, {Azzaro}, {B{\'e}jar}, {Cort{\'e}s-Contreras},
  {Dreizler}, {Hatzes}, {Henning}, {Jeffers}, {Kaminski}, {K{\"u}rster},
  {Montes}, {Morales}, {Oshagh}, {Rodr{\'\i}guez-L{\'o}pez}, {Sch{\"o}fer},
  {Schweitzer}, \& {Zechmeister}}]{Lafarga2021}
{Lafarga}, M., {Ribas}, I., {Reiners}, A., {et~al.} 2021, \aap, 652, A28

\bibitem[{{Lopez-Santiago} {et~al.}(2020){Lopez-Santiago}, {Martino},
  {M{\'\i}guez}, \& {V{\'a}zquez}}]{Lopez-Santiago2020}
{Lopez-Santiago}, J., {Martino}, L., {M{\'\i}guez}, J., \& {V{\'a}zquez}, M.~A.
  2020, \aj, 160, 273

\bibitem[{{Lovis} {et~al.}(2011){Lovis}, {Dumusque}, {Santos}, {Bouchy},
  {Mayor}, {Pepe}, {Queloz}, {S{\'e}gransan}, \& {Udry}}]{Lovis2011}
{Lovis}, C., {Dumusque}, X., {Santos}, N.~C., {et~al.} 2011, arXiv e-prints,
  arXiv:1107.5325

\bibitem[{{Medina} {et~al.}(2022){Medina}, {Charbonneau}, {Winters}, {Irwin},
  \& {Mink}}]{Medina2022}
{Medina}, A.~A., {Charbonneau}, D., {Winters}, J.~G., {Irwin}, J., \& {Mink},
  J. 2022, \apj, 928, 185

\bibitem[{{Mittag} {et~al.}(2019){Mittag}, {Schmitt}, {Hempelmann}, \&
  {Schr{\"o}der}}]{Mittag2019}
{Mittag}, M., {Schmitt}, J.~H.~M.~M., {Hempelmann}, A., \& {Schr{\"o}der},
  K.~P. 2019, \aap, 621, A136

\bibitem[{{Mursula} {et~al.}(2003){Mursula}, {Zieger}, \&
  {Vilppola}}]{Mursula2003}
{Mursula}, K., {Zieger}, B., \& {Vilppola}, J.~H. 2003, \solphys, 212, 201

\bibitem[{{Nagel} {et~al.}(2019){Nagel}, {Czesla}, {Schmitt}, {Dreizler},
  {Anglada-Escud{\'e}}, {Rodr{\'{\i}}guez}, {Ribas}, {Reiners}, {Quirrenbach},
  {Amado}, {Caballero}, {Aceituno}, {B{\'e}jar}, {Cort{\'e}s-Contreras},
  {Gonz{\'a}lez-Cuesta}, {Guenther}, {Henning}, {Jeffers}, {Kaminski},
  {K{\"u}rster}, {Lafarga}, {L{\'o}pez-Gonz{\'a}lez}, {Montes}, {Morales},
  {Passegger}, {Rodr{\'{\i}}guez-L{\'o}pez}, {Schweitzer}, \&
  {Zechmeister}}]{Evangelos}
{Nagel}, E., {Czesla}, S., {Schmitt}, J.~H.~M.~M., {et~al.} 2019, \aap, 622,
  A153

\bibitem[{{Perdelwitz} {et~al.}(2021){Perdelwitz}, {Mittag}, {Tal-Or},
  {Schmitt}, {Caballero}, {Jeffers}, {Reiners}, {Schweitzer}, {Trifonov},
  {Ribas}, {Quirrenbach}, {Amado}, {Seifert}, {Cifuentes},
  {Cort{\'e}s-Contreras}, {Montes}, {Revilla}, \&
  {Skrzypinski}}]{Perdelwitz2021}
{Perdelwitz}, V., {Mittag}, M., {Tal-Or}, L., {et~al.} 2021, \aap, 652, A116

\bibitem[{{Peres} {et~al.}(2000){Peres}, {Orlando}, {Reale}, {Rosner}, \&
  {Hudson}}]{Peres2000}
{Peres}, G., {Orlando}, S., {Reale}, F., {Rosner}, R., \& {Hudson}, H. 2000,
  \apj, 528, 537

\bibitem[{{Queloz} {et~al.}(2001){Queloz}, {Henry}, {Sivan}, {Baliunas},
  {Beuzit}, {Donahue}, {Mayor}, {Naef}, {Perrier}, \& {Udry}}]{Queloz2001}
{Queloz}, D., {Henry}, G.~W., {Sivan}, J.~P., {et~al.} 2001, \aap, 379, 279

\bibitem[{{Quirrenbach} {et~al.}(2020){Quirrenbach}, {CARMENES Consortium},
  {Amado}, {Ribas}, {Reiners}, {Caballero}, {Aceituno}, {Alacid},
  {Alonso-Floriano}, {Anglada-Escud{\'e}}, {Azzaro}, {Baroch}, {Bauer},
  {Becerril}, {B{\'e}jar}, {Bluhm}, {Calvo Ortega}, {Cardona Guill{\'e}n},
  {Casasayas-Barris}, {Chaturvedi}, {Cifuentes}, {Colom{\'e}}, {Conte},
  {Cort{\'e}s-Contreras}, {Czesla}, {D{\'\i}ez-Alonso}, {Dom{\'\i}nguez
  Fern{\'a}ndez}, {Dreizler}, {Duque-Arribas}, {Espinoza}, {Fuhrmeister},
  {Galad{\'\i}-Enr{\'\i}quez}, {Gara Quintana}, {Gonz{\'a}lez-Alvare},
  {Gonz{\'a}lez Cuesta}, {Gonz{\'a}lez Hern{\'a}ndez}, {Guenther}, {de
  Guindos}, {Hatzes}, {Henning}, {Herbort}, {Herrero}, {Hintz},
  {Iglesias-P{\'a}ra}, {Jeffers}, {Johnson}, {de Juan}, {Kaminski}, {Kemmer},
  {Khaimova}, {Khalafinejad}, {Klahr}, {Kossakowski}, {Kreidberg},
  {K{\"u}rster}, {Labarga}, {Lafarga}, {Lamp{\'o}n}, {Lara}, {Lillo-Box},
  {Lodieu}, {L{\'o}pez Gallifa}, {L{\'o}pez Gonz{\'a}lez}, {L{\'o}pez-Puertas},
  {Luque}, {Marfil}, {Mart{\'\i}n-Ruiz}, {Matth{\'e}}, {Molaverdikhani},
  {Montes}, {Morales}, {Morales-Calder{\'o}on}, {Nagel}, {Nortmann}, {Nowak},
  {Ofir}, {Oshaghi}, {Pall{\'e}}, {Passegger}, {Pavlov}, {Pedraz},
  {Perdelwitz}, {Perger}, {Reffert}, {Revilla}, {Rodr{\'\i}guez},
  {Rodr{\'\i}guez L{\'o}pez}, {Sabotta}, {Sadegi}, {Sairam}, {Salz},
  {S{\'a}nchez-L{\'o}pez}, {Sanz-Forcada}, {Sarkis}, {Sch{\"a}fer}, {Schiller},
  {Schlecker}, {Schmitt}, {Sch{\"o}fer}, {Schweitzer}, {Seiferta}, {Shan},
  {Shulyak}, {Skrzypinski}, {Solano}, {Soto}, {Stahl}, {Stangret}, {Stock},
  {Strachan}, {Stuber}, {St{\"u}rmer}, {Tabernero}, {Tal-Or}, {Tala-Pinto},
  {Trifonov}, {Vanaverbeke}, {Yan}, {Zapatero Osorio}, \&
  {Zechmeister}}]{Quirrenbach2020}
{Quirrenbach}, A., {CARMENES Consortium}, {Amado}, P.~J., {et~al.} 2020, in
  Society of Photo-Optical Instrumentation Engineers (SPIE) Conference Series,
  Vol. 11447, Society of Photo-Optical Instrumentation Engineers (SPIE)
  Conference Series, 114473C

\bibitem[{{Reiners} {et~al.}(2018){Reiners}, {Zechmeister}, {Caballero},
  {Ribas}, {Morales}, {Jeffers}, {Sch{\"o}fer}, {Tal-Or}, {Quirrenbach},
  {Amado}, {Kaminski}, {Seifert}, {Abril}, {Aceituno}, {Alonso-Floriano},
  {Ammler-von Eiff}, {Antona}, {Anglada-Escud{\'e}}, {Anwand-Heerwart},
  {Arroyo-Torres}, {Azzaro}, {Baroch}, {Barrado}, {Bauer}, {Becerril},
  {B{\'e}jar}, {Ben{\'{\i}}tez}, {Berdi{\~n}as}, {Bergond}, {Bl{\"u}mcke},
  {Brinkm{\"o}ller}, {del Burgo}, {Cano}, {C{\'a}rdenas V{\'a}zquez}, {Casal},
  {Cifuentes}, {Claret}, {Colom{\'e}}, {Cort{\'e}s-Contreras}, {Czesla},
  {D{\'{\i}}ez-Alonso}, {Dreizler}, {Feiz}, {Fern{\'a}ndez}, {Ferro},
  {Fuhrmeister}, {Galad{\'{\i}}-Enr{\'{\i}}quez}, {Garcia-Piquer},
  {Garc{\'{\i}}a Vargas}, {Gesa}, {G{\'o}mez Galera}, {Gonz{\'a}lez
  Hern{\'a}ndez}, {Gonz{\'a}lez-Peinado}, {Gr{\"o}zinger}, {Grohnert},
  {Gu{\`a}rdia}, {Guenther}, {Guijarro}, {de Guindos}, {Guti{\'e}rrez-Soto},
  {Hagen}, {Hatzes}, {Hauschildt}, {Hedrosa}, {Helmling}, {Henning}, {Hermelo},
  {Hern{\'a}ndez Arab{\'{\i}}}, {Hern{\'a}ndez Casta{\~n}o}, {Hern{\'a}ndez
  Hernando}, {Herrero}, {Huber}, {Huke}, {Johnson}, {de Juan}, {Kim}, {Klein},
  {Kl{\"u}ter}, {Klutsch}, {K{\"u}rster}, {Lafarga}, {Lamert}, {Lamp{\'o}n},
  {Lara}, {Laun}, {Lemke}, {Lenzen}, {Launhardt}, {L{\'o}pez del Fresno},
  {L{\'o}pez-Gonz{\'a}lez}, {L{\'o}pez-Puertas}, {L{\'o}pez Salas},
  {L{\'o}pez-Santiago}, {Luque}, {Mag{\'a}n Madinabeitia}, {Mall}, {Mancini},
  {Mandel}, {Marfil}, {Mar{\'{\i}}n Molina}, {Maroto Fern{\'a}ndez},
  {Mart{\'{\i}}n}, {Mart{\'{\i}}n-Ruiz}, {Marvin}, {Mathar}, {Mirabet},
  {Montes}, {Moreno-Raya}, {Moya}, {Mundt}, {Nagel}, {Naranjo}, {Nortmann},
  {Nowak}, {Ofir}, {Oreiro}, {Pall{\'e}}, {Panduro}, {Pascual}, {Passegger},
  {Pavlov}, {Pedraz}, {P{\'e}rez-Calpena}, {P{\'e}rez Medialdea}, {Perger},
  {Perryman}, {Pluto}, {Rabaza}, {Ram{\'o}n}, {Rebolo}, {Redondo}, {Reffert},
  {Reinhart}, {Rhode}, {Rix}, {Rodler}, {Rodr{\'{\i}}guez},
  {Rodr{\'{\i}}guez-L{\'o}pez}, {Rodr{\'{\i}}guez Trinidad}, {Rohloff},
  {Rosich}, {Sadegi}, {S{\'a}nchez-Blanco}, {S{\'a}nchez Carrasco},
  {S{\'a}nchez-L{\'o}pez}, {Sanz-Forcada}, {Sarkis}, {Sarmiento},
  {Sch{\"a}fer}, {Schmitt}, {Schiller}, {Schweitzer}, {Solano}, {Stahl},
  {Strachan}, {St{\"u}rmer}, {Su{\'a}rez}, {Tabernero}, {Tala}, {Trifonov},
  {Tulloch}, {Ulbrich}, {Veredas}, {Vico Linares}, {Vilardell}, {Wagner},
  {Winkler}, {Wolthoff}, {Xu}, {Yan}, \& {Zapatero Osorio}}]{Reiners2017}
{Reiners}, A., {Zechmeister}, M., {Caballero}, J.~A., {et~al.} 2018, \aap, 612,
  A49

\bibitem[{{Richards} {et~al.}(2009){Richards}, {Rogers}, \&
  {Richards}}]{Richards2009}
{Richards}, M.~T., {Rogers}, M.~L., \& {Richards}, D. S.~P. 2009, \pasp, 121,
  797

\bibitem[{{Roberts} \& {Stix}(1972)}]{Roberts1972}
{Roberts}, P.~H. \& {Stix}, M. 1972, \aap, 18, 453

\bibitem[{{Robertson} {et~al.}(2013){Robertson}, {Endl}, {Cochran}, \&
  {Dodson-Robinson}}]{Robertson2013}
{Robertson}, P., {Endl}, M., {Cochran}, W.~D., \& {Dodson-Robinson}, S.~E.
  2013, \apj, 764, 3

\bibitem[{{Robrade} {et~al.}(2012){Robrade}, {Schmitt}, \&
  {Favata}}]{Robrade2012}
{Robrade}, J., {Schmitt}, J.~H.~M.~M., \& {Favata}, F. 2012, \aap, 543, A84

\bibitem[{{Sch{\"o}fer} {et~al.}(2019){Sch{\"o}fer}, {Jeffers}, {Reiners},
  {Shulyak}, {Fuhrmeister}, {Johnson}, {Zechmeister}, {Ribas}, {Quirrenbach},
  {Amado}, {Caballero}, {Anglada-Escud{\'e}}, {Bauer}, {B{\'e}jar},
  {Cort{\'e}s-Contreras}, {Dreizler}, {Guenther}, {Kaminski}, {K{\"u}rster},
  {Lafarga}, {Montes}, {Morales}, {Pedraz}, \& {Tal-Or}}]{Patrick}
{Sch{\"o}fer}, P., {Jeffers}, S.~V., {Reiners}, A., {et~al.} 2019, \aap, 623,
  A44

\bibitem[{{Schweitzer} {et~al.}(2019){Schweitzer}, {Passegger}, {Cifuentes},
  {B{\'e}jar}, {Cort{\'e}s-Contreras}, {Caballero}, {del Burgo}, {Czesla},
  {K{\"u}rster}, {Montes}, {Zapatero Osorio}, {Ribas}, {Reiners},
  {Quirrenbach}, {Amado}, {Aceituno}, {Anglada-Escud{\'e}}, {Bauer},
  {Dreizler}, {Jeffers}, {Guenther}, {Henning}, {Kaminski}, {Lafarga},
  {Marfil}, {Morales}, {Schmitt}, {Seifert}, {Solano}, {Tabernero}, \&
  {Zechmeister}}]{Schweitzer2019}
{Schweitzer}, A., {Passegger}, V.~M., {Cifuentes}, C., {et~al.} 2019, \aap,
  625, A68

\bibitem[{{Stock} {et~al.}(2020){Stock}, {Nagel}, {Kemmer}, {Passegger},
  {Reffert}, {Quirrenbach}, {Caballero}, {Czesla}, {B{\'e}jar}, {Cardona},
  {D{\'\i}ez-Alonso}, {Herrero}, {Lalitha}, {Schlecker}, {Tal-Or},
  {Rodr{\'\i}guez}, {Rodr{\'\i}guez-L{\'o}pez}, {Ribas}, {Reiners}, {Amado},
  {Bauer}, {Bluhm}, {Cort{\'e}s-Contreras}, {Gonz{\'a}lez-Cuesta}, {Dreizler},
  {Hatzes}, {Henning}, {Jeffers}, {Kaminski}, {K{\"u}rster}, {Lafarga},
  {L{\'o}pez-Gonz{\'a}lez}, {Montes}, {Morales}, {Pedraz}, {Sch{\"o}fer},
  {Schweitzer}, {Trifonov}, {Zapatero Osorio}, \& {Zechmeister}}]{SN20}
{Stock}, S., {Nagel}, E., {Kemmer}, J., {et~al.} 2020, \aap, 643, A112

\bibitem[{{Su{\'a}rez Mascare{\~n}o} {et~al.}(2016){Su{\'a}rez Mascare{\~n}o},
  {Rebolo}, \& {Gonz{\'a}lez Hern{\'a}ndez}}]{Suarez-Mascareno2016}
{Su{\'a}rez Mascare{\~n}o}, A., {Rebolo}, R., \& {Gonz{\'a}lez Hern{\'a}ndez},
  J.~I. 2016, \aap, 595, A12

\bibitem[{{Su{\'a}rez Mascare{\~n}o} {et~al.}(2018){Su{\'a}rez Mascare{\~n}o},
  {Rebolo}, {Gonz{\'a}lez Hern{\'a}ndez}, {Toledo-Padr{\'o}n}, {Perger},
  {Ribas}, {Affer}, {Micela}, {Damasso}, {Maldonado}, {Gonz{\'a}lez-Alvarez},
  {Leto}, {Pagano}, {Scandariato}, {Sozzetti}, {Lanza}, {Malavolta}, {Claudi},
  {Cosentino}, {Desidera}, {Giacobbe}, {Maggio}, {Rainer}, {Esposito},
  {Benatti}, {Pedani}, {Morales}, {Herrero}, {Lafarga}, {Rosich}, \&
  {Pinamonti}}]{SuarezMascareno2018}
{Su{\'a}rez Mascare{\~n}o}, A., {Rebolo}, R., {Gonz{\'a}lez Hern{\'a}ndez},
  J.~I., {et~al.} 2018, \aap, 612, A89

\bibitem[{{Wargelin} {et~al.}(2017){Wargelin}, {Saar}, {Pojma{\'n}ski},
  {Drake}, \& {Kashyap}}]{Wargelin2017}
{Wargelin}, B.~J., {Saar}, S.~H., {Pojma{\'n}ski}, G., {Drake}, J.~J., \&
  {Kashyap}, V.~L. 2017, \mnras, 464, 3281

\bibitem[{{Wilson}(1978)}]{Wilson1978}
{Wilson}, O.~C. 1978, \apj, 226, 379

\bibitem[{{Yadav} {et~al.}(2016){Yadav}, {Christensen}, {Wolk}, \&
  {Poppenhaeger}}]{Yadav2016}
{Yadav}, R.~K., {Christensen}, U.~R., {Wolk}, S.~J., \& {Poppenhaeger}, K.
  2016, \apjl, 833, L28

\bibitem[{{Zechmeister} {et~al.}(2014){Zechmeister}, {Anglada-Escud{\'e}}, \&
  {Reiners}}]{pipeline}
{Zechmeister}, M., {Anglada-Escud{\'e}}, G., \& {Reiners}, A. 2014, \aap, 561,
  A59

\bibitem[{{Zechmeister} \& {K{\"u}rster}(2009)}]{Zechmeister2009}
{Zechmeister}, M. \& {K{\"u}rster}, M. 2009, \aap, 496, 577

\bibitem[{{Zechmeister} {et~al.}(2018){Zechmeister}, {Reiners}, {Amado},
  {Azzaro}, {Bauer}, {B{\'e}jar}, {Caballero}, {Guenther}, {Hagen}, {Jeffers},
  {Kaminski}, {K{\"u}rster}, {Launhardt}, {Montes}, {Morales}, {Quirrenbach},
  {Reffert}, {Ribas}, {Seifert}, {Tal-Or}, \& {Wolthoff}}]{serval}
{Zechmeister}, M., {Reiners}, A., {Amado}, P.~J., {et~al.} 2018, \aap, 609, A12

\end{thebibliography}

\appendix
\section{Further examples of long-term variability}\label{app:a}



\begin{figure}
\begin{center}
\includegraphics[width=0.5\textwidth, clip]{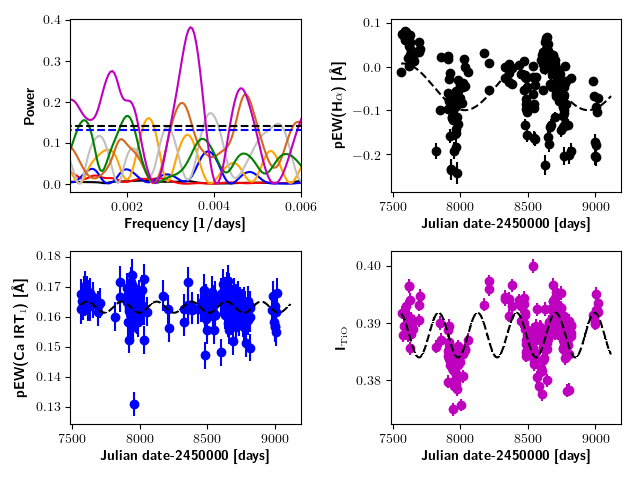}
        \caption{\label{fig:M5} Same as in Fig.~\ref{fig:J22330}. GLS and time series for J20260+585 / Wolf~1069.
        In pEW(H$\alpha$), frequent flaring and the spacing of the observation leads
        to the period we found, while the same period may be present in the
        I$_{\rm TiO}$ data.}
\end{center}
\end{figure}

\begin{figure}
\begin{center}
\includegraphics[width=0.5\textwidth, clip]{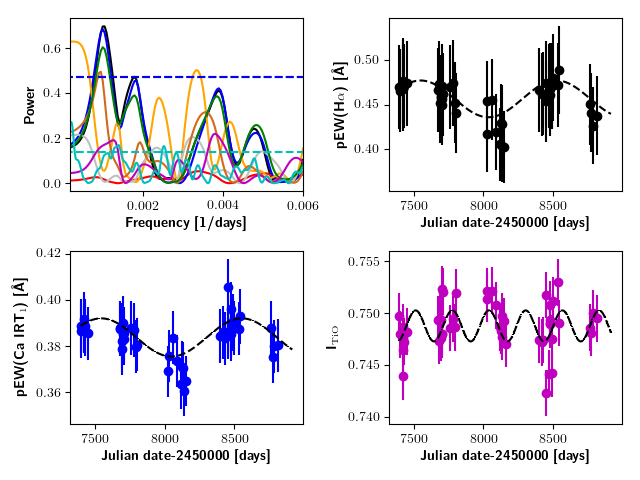}
\caption{\label{fig:J06105} Same as in Fig.~\ref{fig:J22330}, but for J06105$-$218 / HD~42581~A. 
        }
\end{center}
\end{figure}

\begin{figure}
\begin{center}
\includegraphics[width=0.5\textwidth, clip]{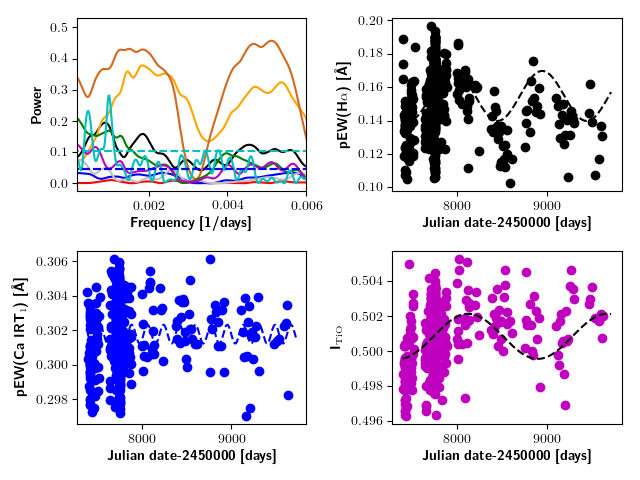}
\caption{\label{fig:J07274} Same as in Fig.~\ref{fig:J22330}, but for J07274+052 / Luyten's star.
  We omit the error bars for clarity.
        }
\end{center}
\end{figure}


\begin{figure*}
\begin{center}
\includegraphics[width=0.5\textwidth, clip]{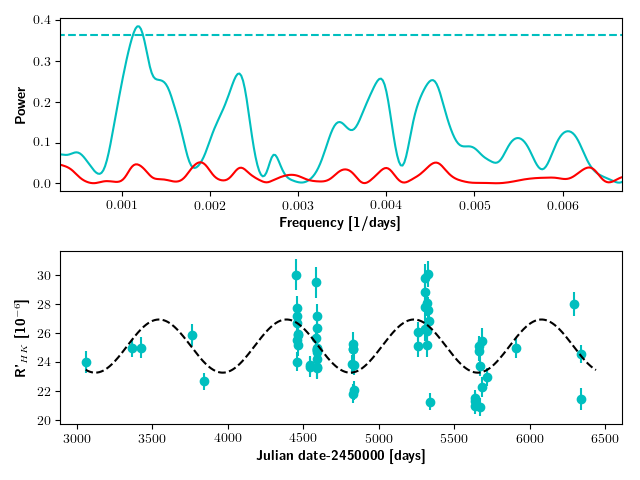}
\includegraphics[width=0.5\textwidth, clip]{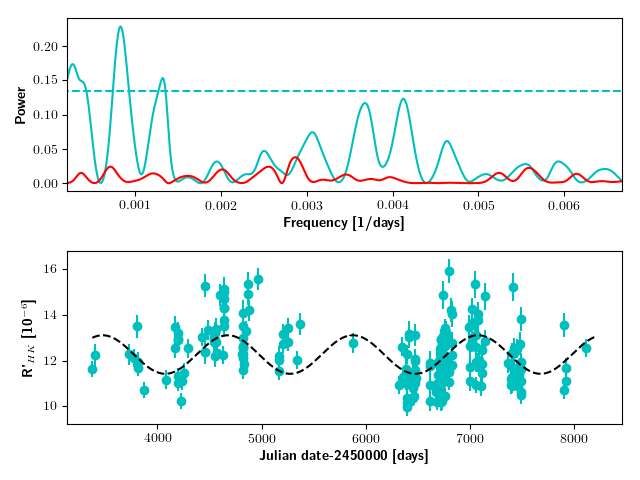}
\caption{\label{fig:rpalone1} GLS and time series of J10251$-$102 / BD$-$09~3070 (M1.0\,V, \emph{left}) and
  J10289+008 / BD+01~2447 (M2.0\,V, \emph{right}). In the GLS (\emph{top}), we show the power of the \rpr\
  data in cyan, the dashed cyan line marks the 0.0005 FAP level, and the red line
  is the power of the window function. In the time series (\emph{bottom}), we show the
  \rpr\ values as cyan dots and the best-fit sine function as the dashed black line.
}
\end{center}
\end{figure*}

\begin{figure*}
\begin{center}
\includegraphics[width=0.5\textwidth, clip]{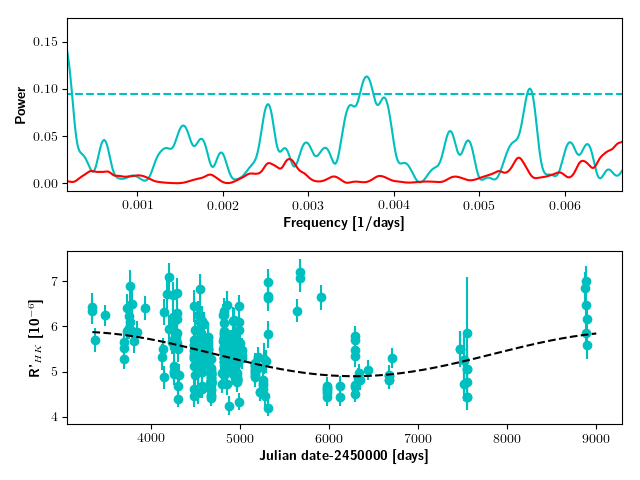}
\includegraphics[width=0.5\textwidth, clip]{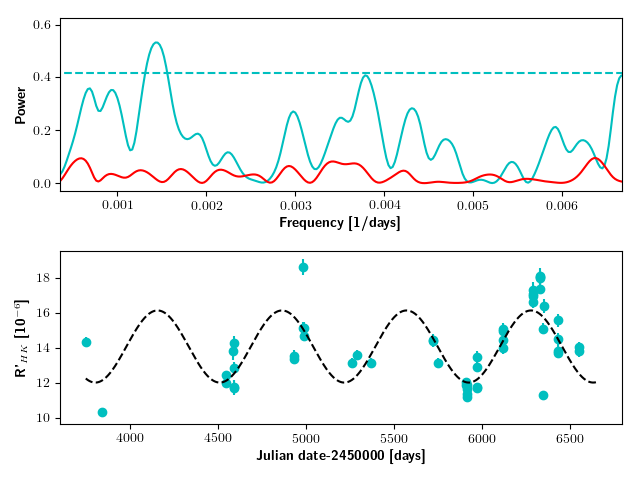}
\caption{\label{fig:rpalone2} Same as Fig. \ref{fig:rpalone1}, but for  
  J11421+267 / Ross~905 (M2.5\,V, \emph{left}) and J14294+155 / Ross130 (M2.0\,V, \emph{right}). 
}
\end{center}
\end{figure*}

\begin{figure*}
\begin{center}
\includegraphics[width=0.5\textwidth, clip]{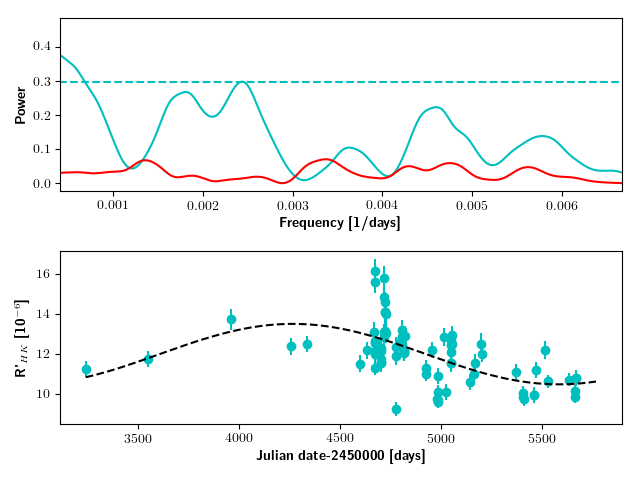}
\includegraphics[width=0.5\textwidth, clip]{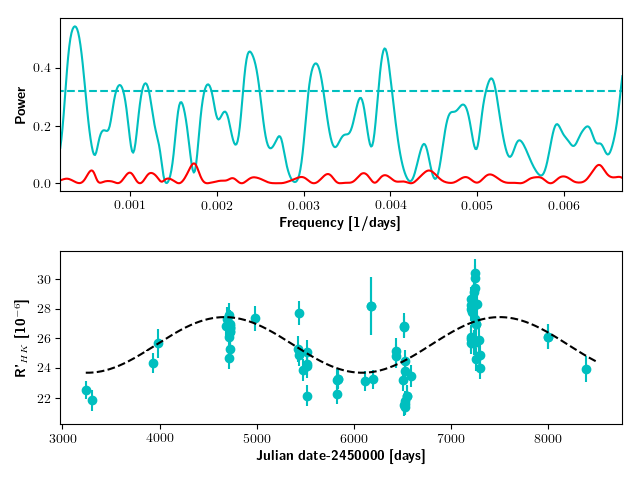}
\caption{\label{fig:rpalone3} Same as Fig. \ref{fig:rpalone1}, but for
  J20450+444 / BD+44~3567 (M1.5\,V, \emph{left}) and J20533+621 / HD~199305 (M1.0\,V, \emph{right}). 
}
\end{center}
\end{figure*}


\end{document}